\documentclass[12pt, draftclsnofoot, onecolumn]{IEEEtran}

\usepackage[pdftex]{graphicx}
\usepackage[cmex10]{amsmath,empheq}
\usepackage{cite}
\usepackage{url}
\usepackage{color}
\usepackage{microtype}
\usepackage{amssymb}
\usepackage{relsize}
\usepackage{lipsum}
\usepackage{epstopdf}
\usepackage{enumitem}
\usepackage{bm}
\usepackage{nicefrac}
\usepackage[percent]{overpic}
\usepackage{subfig}

\begin{document}

\title{DeepRx: Fully Convolutional Deep Learning Receiver}

\newcommand{\todo}[1]{}
\newcommand{\changed}[1]{{\color{blue}{#1}}}
\renewcommand{\todo}[1]{{\color{red}{#1}}\PackageWarning{TODO:}{#1!}}

\author{Mikko~Honkala,
        Dani~Korpi,
        and~Janne~M.J. Huttunen
\thanks{M. Honkala, D. Korpi, and J. Huttunen are with Nokia Bell Labs, Espoo, Finland.}
}

\maketitle

\begin{abstract}

Deep learning has solved many problems that are out of reach of heuristic algorithms. It has also been successfully applied in wireless communications, even though the current radio systems are well-understood and optimal algorithms exist for many tasks.
While some gains have been obtained by learning individual parts of a receiver, a better approach is to jointly learn the whole receiver. This, however, often results in a challenging nonlinear problem, for which the optimal solution is infeasible to implement.
To this end, we propose a deep fully convolutional neural network, DeepRx, which executes the whole receiver pipeline from frequency domain signal stream to uncoded bits in a 5G-compliant fashion. We facilitate accurate channel estimation by constructing the input of the convolutional neural network in a very specific manner using both the data and pilot symbols. Also, DeepRx outputs soft bits that are compatible with the channel coding used in 5G systems.
Using 3GPP-defined channel models, we demonstrate that DeepRx outperforms traditional methods. We also show that the high performance can likely be attributed to DeepRx learning to utilize the known constellation points of the unknown data symbols, together with the local symbol distribution, for improved detection accuracy.

\end{abstract}


\section{Introduction}

The recent advances in deep learning techniques have resulted in new applications of neural networks in various fields, including wireless communications \cite{Zhang19a,Jiang17a,oshea17,Fu18a,zhao2018,neumann18}. Machine learning (ML) has already been shown to be very effective in optimizing the higher layers of the communication stack \cite{Morocho19a,Chinchali18a,Tang18a,Xu17a}. However, it is indisputable that the foundation of the overall network-level performance is set by the processing employed in the physical layer. Therefore, in this article, we show that there are unrealized gains to be achieved also in the physical layer processing via the use of ML, with which we can improve the radio performance of the individual devices within the network.

In particular, by treating the radio receiver implementation as one supervised learning problem, it is possible to consider many of the individual receiver tasks, such as channel estimation and equalization, jointly. The thesis of this work is that this will result in a higher performance than optimizing each individual component separately, since the optimization target can closely mirror the real-world target, meaning that the resulting model is not bound by unrealistic or inaccurate assumptions. In this article, our approach is to train a deep neural network to detect the received bits from the received waveform. Namely, we consider the physical layer processing of a 5G-compliant radio receiver, whose task is to obtain the information bits from an orthogonal frequency-division multiplexing (OFDM) waveform, modulated in accordance with 5G numerology \cite{NR_38211}. The benefit of this type of an approach is that the receiver's task can be represented as a supervised learning problem without requiring any labeling by existing algorithms or by humans. Indeed, the input data is simply the received waveform in the frequency domain, while the original transmitted bits are the corresponding labels.

There are already several studies that propose implementing certain parts of the digital receiver chain using a neural network. For instance, \cite{neumann18} considers a neural network-aided channel estimator, which is shown to approach the performance of a genie-aided estimator when the number of RX antennas is large, while the ML-based mmWave channel estimation solution proposed in \cite{he18} is shown to outperform existing compressed sensing-based schemes throughout the signal-to-noise ratio (SNR) range. In \cite{chang19} Chang et al. apply convolutional neural networks (CNNs) \cite{lecun98, Bengio09, LeCun15} to equalization, achieving a lower error vector magnitude than that of multi-modulus algorithm or recursive least squares-based approaches. Deep learning-based demapping is analyzed in \cite{shental2019}, where Shental and Hoydis propose a deep neural network approach for efficiently calculating bit log-likelihood ratios (LLRs) of equalized symbols. The proposed deep learning based demapper is shown to achieve similar accuracy as the optimal log maximum a-posteriori rule, albeit with greatly reduced computational cost. In addition, there are some works that propose augmenting traditional receiver processing flow with deep learning components to improve the performance \cite{gao18,He19a,Samuel19a}, each of them outperforming the traditional receiver benchmark, as long as proper training is conducted.

The prospect of implementing larger portions of the receiver using a single neural network has also been considered by some authors. For instance, Ye et al. \cite{ye18} investigate combined channel estimation and signal detection using deep learning. There, the detection is carried out using a fully-connected neural network that processes the pilots and the data signal. Such a fully learned receiver is shown to clearly outperform a minimum mean square error (MMSE) based traditional receiver when there are few channel estimation pilots or when the cyclic prefix is omitted. In \cite{zhao2018}, on the other hand, CNNs are applied to implement a receiver that extracts the bit estimates directly from a time-domain RX signal, achieving excellent performance at low-to-mid SNRs. At high SNRs, the CNN-based scheme still outperforms a linear least squares-based receiver, while falling behind an MMSE-based receiver and a genie-aided receiver with perfect channel knowledge. As the most extreme case, deep learning-based end-to-end solutions, where both the transmitter and receiver are learned simultaneously from the data without any prespecified modulation scheme or waveform, have also been widely studied \cite{oshea17,dorner18, aoudia2018,cammerer2019}. Such schemes have been shown to have potential to outperform traditional heuristic radio links, e.g., by learning a better constellation shape \cite{cammerer2019}.

Many of these prior works have successfully implemented a neural-network-based radio receiver and have also demonstrated high performance in comparison to the traditional receiver algorithms. These findings indicate that developing a data-driven receiver with deep neural network can indeed increase the performance of the future radio systems. However, we show that by carefully designing the neural network architecture and its inputs, it is possible to achieve an even higher increase in performance. Our findings indicate that most gains are obtained by allowing the neural network to utilize the unknown data symbols and their distribution in enhancing the channel estimation accuracy.

In contrast to many related works, we also consider standards compliance, in particular with 5G New Radio (NR). In order for the learned receivers to achieve 5G compliance, they must support, among other things, the various different demodulation reference signal (DMRS) or pilot configurations included in the NR specifications. In addition, the output of the receiver must be decoded by a low-density parity-check (LDPC) decoder, which means that the neural network must be capable of estimating also the uncertainty of each received bit for each modulation order. Consequently, there is a need for flexible deep learning-based receiver solutions that can handle all the different reference signal configurations and modulation schemes within a single implementation while being compliant with the other processing stages. 

In this article, we propose a fully convolutional neural network architecture, referred to as DeepRx, that learns a high-performance OFDM receiver from the data. This is achieved by feeding a frequency domain representation of the received signal over the whole transmission time interval (TTI) to a CNN, and training it to output the LLRs of the transmitted bits (soft bits) without restricting the individual processing stages in any way. Furthermore, the CNN-based DeepRx is trained to operate under various different scenarios, parameters and configurations. Therefore, with the proposed approach, it suffices to implement only the transmitter manually; the receiver can be learned based on the received waveform and the known transmitted bits. In this work, training and validation of the proposed architecture is carried out using simulated uplink (UL) data, generated with Matlab's 5G toolbox \cite{Matlab5G}. The performance of the different receivers is evaluated using the bit error rate (BER) both before and after LDPC decoding.

In particular, the main contributions of this article are as follows:
\begin{itemize}
	\item We describe a novel deep learning receiver (DeepRx), which outperforms traditional receivers in terms of radio performance and is among the best reported in literature. The main reasons for the performance increase are:
			a) DeepRx is trained to obtain LLRs directly from frequency domain antenna data, which allows it to carry out all tasks (channel estimation, equalization and soft demapping) jointly,
			b) it is allowed to utilize, in addition to pilots, the received data symbols and their distributions during all the tasks, which is especially helpful under high Doppler shifts and sparse pilot configurations,
			c) the utilized training approach allows the receiver to cope well also with non-Gaussian noise, such as interference from another cell.
	\item We design the DeepRx network to be 5G-compliant in several aspects and show that it is possible for a single network to cover large parts of the 5G specification efficiently. The same network can operate under arbitrary DMRS configurations, modulation orders, and code rates\footnote{It is to be noted that, even though particular attention is paid to 5G compliance, the proposed CNN architecture is compatible with any communication system employing OFDM waveforms, such as WiFi.}. We also demonstrate that the LLRs provided by the DeepRx network can be processed by a 5G-compliant LDPC decoder, resulting in state of the art performance.
	\item We provide extensive performance comparisons demonstrating that DeepRx can outperform the benchmark receiver algorithms. We also provide example results with carefully manipulated input signals to gain insight into its behavior and outstanding performance.
\end{itemize}

The rest of this paper is organized as follows. In Section~\ref{sec:simulator}, we describe the reference transmitter and receiver architectures which are used as a basis for generating training and validation data. We also give a brief overview into traditional receiver architectures that are used as baselines in our study. Then, Section~\ref{sec:nn_rx} discusses in detail the proposed DeepRx architecture and the training procedure. After this, Section~\ref{sec:training_data} outlines the simulation-based training and validation data generation, while Section~\ref{sec:validation_res} shows the validation results with comparisons against baseline receivers. Finally, Section~\ref{sec:conc} concludes the article.

\section{System Model} 
\label{sec:simulator}

\begin{figure*}[t!]
	\centering
	\includegraphics[width=\textwidth]{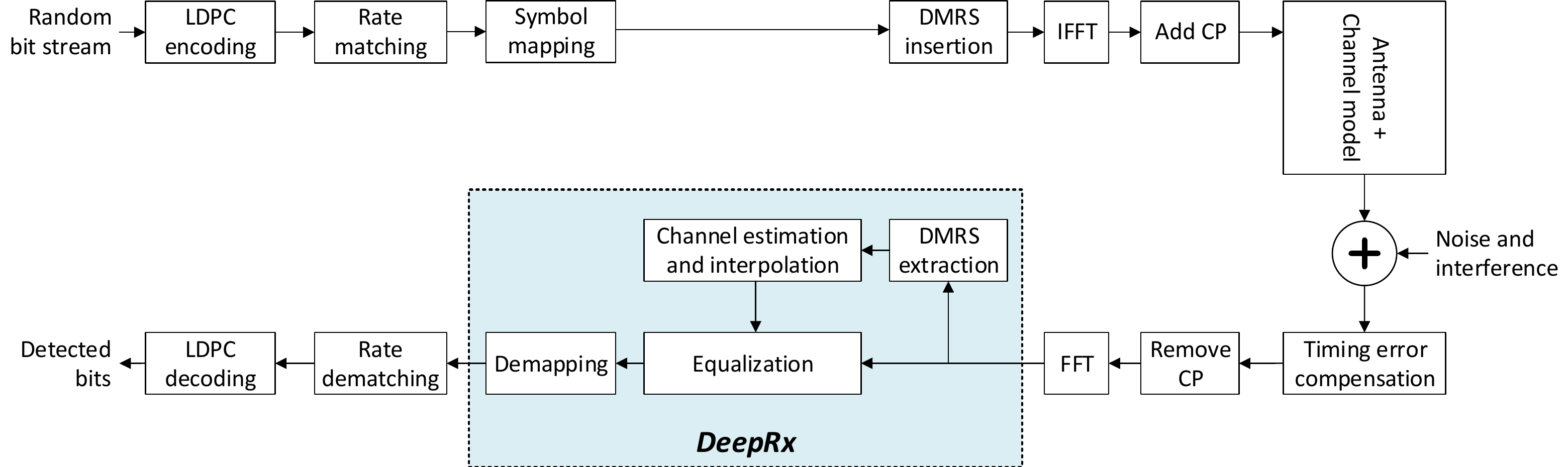}
	\caption{Illustration of the link-level simulator used for data collection and benchmarking.}
	\label{fig:simulator}
\end{figure*}

For data generation, we consider a 5G-compliant physical uplink shared channel (PUSCH) simulator, implemented with MATLAB's 5G Toolbox \cite{Matlab5G}. The simulator architecture is illustrated in Fig.~\ref{fig:simulator} and it includes all the physical layer components, starting from the transport block (TB) information bits, and terminating after the decoding phase. The hybrid automatic repeat request (HARQ) procedure is omitted from the analysis of this article since the proposed DeepRx architecture is completely transparent to the processes of the higher layers. Moreover, in this study, we restrict the number of transmit antennas to one, considering a single-input and multiple-output (SIMO) system. Extending the work to a multiple-input and multiple-output (MIMO) system is left as a future work item.

As a starting point, a specified amount of uniformly distributed information bits are randomly generated. These are then encoded with an LDPC code and fed through the rate matching processing. The resulting code word is mapped into symbols, and these symbols are distributed over the available physical resource blocks (PRBs) within the TTI. Demodulation reference signals (DMRS), or pilots, are also inserted into the specified subcarriers. After this, the data is turned into an OFDM waveform by feeding the PRBs into an inverse Fourier transform (IFFT), resulting in 14 individual OFDM symbols per each TTI. Before transmission, a cyclic prefix (CP) is added to the beginning of each OFDM symbol to mitigate inter-symbol-interference.

Having obtained the transmit waveform for the whole TTI, it is fed through a channel model. For the purposes of this work, we utilize ten different channel models specified by 3GPP \cite{NR_38901}. They include five different clustered delay line (CDL) and tapped delay line (TDL) channel models, each with their own delay profile. Out of these ten models, four represent a non-line-of-sight (NLOS) scenario, while the remaining are line-of-sight (LOS) channel models. 
The channel model is chosen randomly for each frame. In addition, the maximum Doppler shift and root mean square (RMS) delay spread are also randomly chosen for each individual channel realization.

After propagating through the channel, the received waveform is subjected to observation noise (Gaussian white noise is assumed) and interference. In this article, the SNR is defined over the whole band, meaning that the SNR upon detection is somewhat higher since some of the subcarriers are unused.

Some of the studied experiments also include inter-cell-interference. The interference is represented as another waveform with similar numerology but different random information bits. Moreover, it is also fed through a different random channel realization and has a random time offset with respect to the desired signal. The power of the interference with respect to the desired signal (signal-to-interference ratio, SIR) is randomly chosen.

Having added the noise and interference, the total received waveform is fed to the receiver for detecting the information bits. The first stage in the receiver is timing offset estimation of the received waveform, which is done based on the known channel delay plus a random offset to model practical timing errors. After this, the signal is demodulated, which simply consists of removing the CP and calculating the fast Fourier transform (FFT) of each individual OFDM symbol. In this case, each TTI consists of 14 OFDM symbols, as already mentioned above. The received signal after the FFT can be expressed as
\begin{align}
\mathbf{y}_{ij} = \mathbf{H}_{ij} x_{ij} + \mathbf{n}_{ij}, \label{eq:rx_signal}
\end{align}
where $i$ and $j$ denote the OFDM symbol and subcarrier indices, respectively, $\mathbf{y}_{ij} \in \mathbb{C}^{N_r \times 1}$ and $x_{ij} \in \mathbb{C}$ are the received and transmitted symbols, respectively, $\mathbf{H}_{ij} \in \mathbb{C}^{N_r \times 1}$ is the channel in the $i$th OFDM symbol over the $j$th subcarrier, $\mathbf{n}_{ij} \in \mathbb{C}^{N_r \times 1}$ is the noise-plus-interference signal, and $N_r$ is the number of RX antennas. The total number of OFDM symbols is denoted by $S$ (which in the context of 5G is fixed at $S = 14$, as mentioned above), while the total amount of subcarriers is denoted by $F$, meaning that $i =0,\hdots,S-1$ and $j = 0,\hdots,F-1$.

The frequency-domain samples of the received 14 OFDM symbols, represented in \eqref{eq:rx_signal}, constitute the input of the actual receiver processing. In the following section, we explain how they are traditionally processed using least squares (LS) channel estimation and linear minimum mean square error (LMMSE) equalization (our baseline), while in Section~\ref{sec:nn_rx} we describe how the processing can be carried out by a deep CNN (DeepRx).

\subsection*{Traditional Receiver Processing}
\label{sec:trad-rece-proc}

The first step in OFDM receiver processing is to estimate the channel using the known pilots (DMRS) and assuming frequency-flat fading for the individual subcarriers. More precisely, using~\eqref{eq:rx_signal}, the raw channel estimate is first calculated as (note that the pilots lie on the unit circle in the complex domain):
\begin{align}
  \label{eq:raw_channel_estimate}
\widehat{\mathbf{H}}_{ij} = \mathbf{y}_{ij} x_{ij}^*, \quad (i,j) \in \mathcal{P}
\end{align}
where $\mathcal{P}$ denotes the set of indices corresponding to pilot locations in the time-frequency grid, and $(\cdot)^*$ denotes the complex conjugate. The raw channel estimate is then interpolated to fill the whole time-frequency grid and thereby provide channel estimates for the data symbols. This will result in the channel estimate $\widehat{\mathbf{H}}_{ij} \in \mathbb{C}^{N_r \times 1}$ for $(i,j) \in \mathcal{D}$, where $\mathcal{D}$ denotes the set of indices of the data symbols and subcarriers. Also the noise(-plus-interference) power $\sigma^2_n$ is estimated during the channel estimation phase.

Each data symbol is then equalized using the interpolated channel estimate. As mentioned above, an LMMSE equalizer is used in the considered reference receiver architecture, which means that the equalizer output for $(i,j) \in \mathcal{D}$ is given by
\begin{align}
\hat{x}_{ij} = \left(\widehat{\mathbf{H}}_{ij}^H \widehat{\mathbf{H}}_{ij} + \hat{\sigma}^2_n \mathbf{I}\right)^{-1} \widehat{\mathbf{H}}_{ij}^H \mathbf{y}_{ij}, \label{eq:mmse_eq}
\end{align}
where $\hat{\sigma}^2_n$ is the noise power estimate, $\mathbf{I}$ is the identity matrix and $(\cdot)^H$ denotes the Hermitian transpose. 

The equalized symbols are next fed to the demapper, which calculates the soft bits or LLRs based on the symbol estimates $\hat{x}_{ij}$. The LLRs are defined by  
\begin{align}
	L_{ijl} \triangleq \log \left( \frac{\operatorname{Pr}\left( c_l = 0 | \hat{x}_{ij}\right)}{\operatorname{Pr}\left( c_l = 1 | \hat{x}_{ij}\right)} \right), \label{eq:llr_exact}
\end{align}
where $\operatorname{Pr}\left( c_l = b | \hat{x}_{ij}\right)$ is the conditional probability that the transmitted bit $c_l$ is $b \in \{0,1\}$ given the observed symbol $\hat{x}_{ij}$, and $l = 0,\hdots,B-1$ where $B$ is the number of bits per symbol. Assuming that the equalizer removes all the channel effects and only Gaussian white noise remains, the LLRs can be approximated with good accuracy by
\begin{align}
	L_{ijl} \approx \frac{1}{\hat{\sigma}_n^2} \left( \min_{x \in C_l^1} \left\| \hat{x}_{ij} - x \right\|_2^2 - \min_{x \in C_l^0} \left\| \hat{x}_{ij} - x \right\|_2^2\right), \label{eq:llr_appr}
\end{align}
where $x \in C_l^b$ represent those points in constellation $C$ for which the $l$th bit is $b \in \{0,1\}$, and $\hat{\sigma}_n^2$ is again the noise power estimate. In the reference implementation the final LLRs are also scaled by the channel magnitude of the considered subcarrier to reflect the higher uncertainty due to more severe fading.

\section{Convolutional Neural Network-Based Receiver}
\label{sec:nn_rx} 

Here we turn the focus to the design rationale of the DeepRx network architecture. DeepRx operates on the Fourier transformed frequency-domain data (see Fig. \ref{fig:simulator}) collected during a TTI and its output are the final bit-level LLRs. Inputting the whole TTI at once allows the network to utilize all the information therein for estimating each of the bits. Moreover, given a non-static environment and potentially mobile UEs, the frequency-domain channel coefficients are different for each subcarrier and OFDM symbol. Considering that the physical channels in such cases are locally strongly correlated in frequency and in time, we employ a fully-convolutional neural network, where 2D convolutions operate in frequency and time dimensions. The objective of these 2D CNN filters is to learn such local correlations that are not frequency or time dependent, and re-use them effectively over the whole TTI.

Another design rationale is that, as the sparse pilot symbols only provide local information of the channel, we allow the network to utilize the unknown data and its known distribution for improved estimation of LLRs far away from the actual pilot locations. Therefore, we give the CNN unrestricted access to all data instead of designing separate blocks or paths for pilot-based channel estimation and data symbol equalization as in most related works. We assume that the channel and LLR estimation can be improved if the whole TTI (both the unknown received data and the known pilots) is inputted to the network in a coherent fashion, since this allows it to utilize all the data to carry out the assigned task.

While this design principle can also be extended to the case where several signal streams are spatially multiplexed in MIMO operation, the forthcoming description is written for a SIMO case where there is just one signal present. Extending the DeepRx architecture to full MIMO processing is an important future work item for us.

\begin{figure}[ht!]
	\centering
	\includegraphics[width=0.5\linewidth]{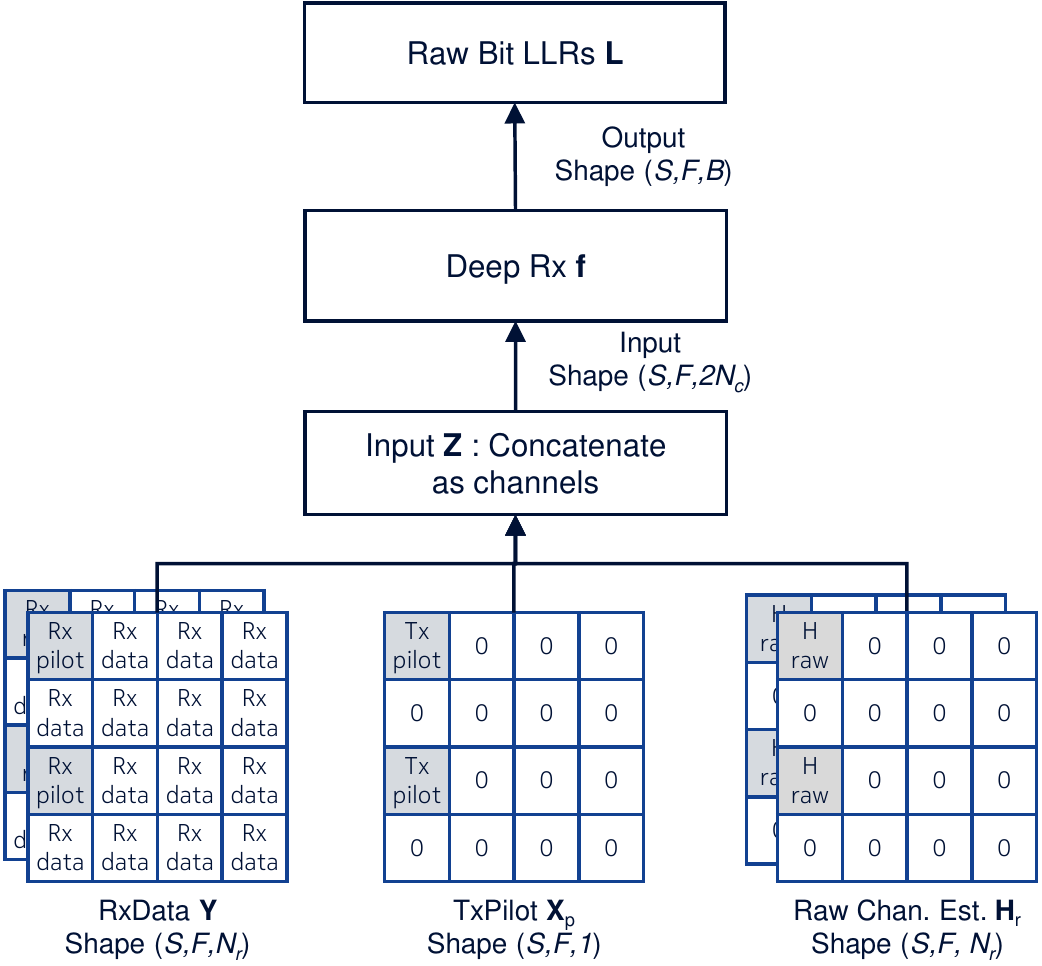}
	\caption{Input to the DeepRx is a concatenation of the received unknown data, the known pilot symbols and the raw channel estimates.}
	\label{fig:dl_input}
\end{figure}

As shown in Fig.~\ref{fig:dl_input}, DeepRx operates on a three-dimensional input array consisting of received data and pilot information, which is constructed as follows. 
\begin{itemize}
\item The first part of input is the received signal after the FFT, denoted by $\mathbf{Y}\in\mathbb{C}^{S \times F \times N_r}$, which contains both data and received pilot symbols (recall that $S$ is number of symbols in time, $F$ is number of subcarriers, and $N_r$ is the number of RX antennas).
\item The second part is $\mathbf{X}_p\in\mathbb{C}^{S \times F}$ which contains the pilot reference symbols positioned so that they correspond to the pilot positions within the received signal $\mathbf{Y}$ in both frequency and time, the non-pilot positions being filled with zeros (see Fig. \ref{fig:dl_input} for illustration).
\item In addition, we pre-compute the raw channel estimate $\mathbf{\widehat{H}}_r = \mathbf{Y} \odot \mathbf{X}_p^*$ for the pilot positions, where $\odot$ and $(\cdot)^*$ are the element wise product and complex conjugate, respectively, and give this as the third part of the input. In the case where there are multiple RX antennas, the elements of $\mathbf{X}_p$ are duplicated along the third dimension when carrying out the raw channel estimation.
\end{itemize}
 
As the first two dimensions of $\mathbf{Y}$, $\mathbf{X}_p$ and $\mathbf{\widehat{H}}_r$ are equal, they can be stacked together along the third dimension (channel) to form $\mathbf{Z}_c \in\mathbb{C}^{S \times F \times N_c}$ where $N_c = 2N_r + 1$. Furthermore, we convert the complex-valued input into real-valued by stacking the real and imaginary parts of the input as separate channels, resulting in the final input array $\mathbf{Z}\in\mathbb{R}^{S \times F \times 2N_c}$. Stacking the relevant data into channels in this manner allows the convolutions to operate on data that is mutually related. While it is also possible to use a complex-valued network with a complex-valued input, we have not observed any performance gains in doing so. 

 {\bf Remark:} The input $\mathbf{X}_p$ includes all required information about the pilot positions, which can have various different configurations. Via this, our setup facilitates a single network to operate with multiple pilot configurations, given that they have been presented to the network during training. We have experimentally verified this by training a single network that successfully handled all the pilot configurations defined in the 5G standard. It should also be noted that while it is not strictly necessary to feed the raw-channel estimate $\mathbf{\widehat{H}}_r$ as input to DeepRx, doing so allows for a somewhat easier and faster learning process (approximating such operation with a neural network can require several layers).
 
\begin{figure}[ht!]
\centering
\includegraphics[width=0.2\linewidth]{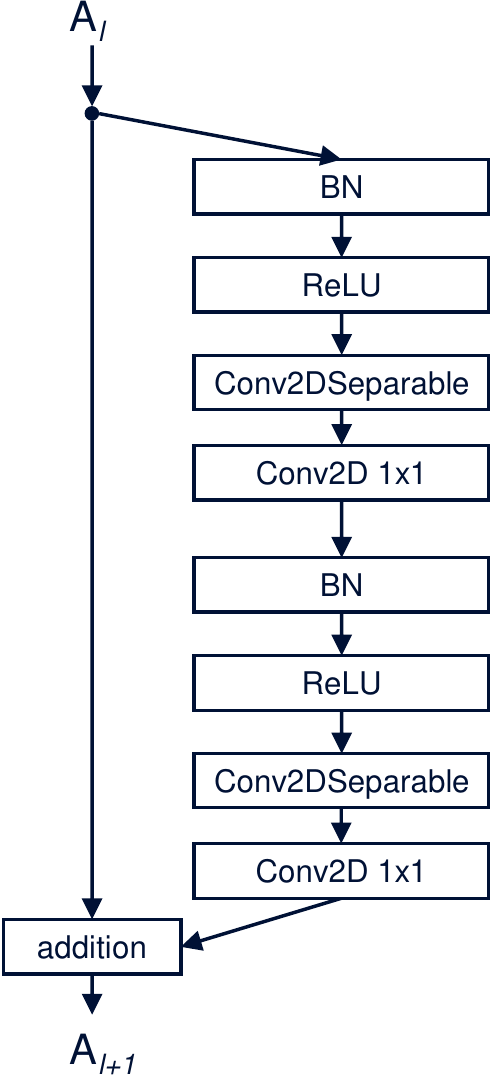}
\caption{Preactivation ResNet Block between previous layer $A_l$ and the next layer $A_l+1$. Blocks ``Conv2DSeparable'' and ``Conv2D 1x1'' form a depthwise separable convolution introduced in \cite{conf/cvpr/SifreM13,chollet2017}. BN and ReLU stand for batch normalization and rectified linear unit, respectively.} \label{fig:resnet}
\end{figure}

 For the neural network $\mathbf{f}: \mathbf{Z} \rightarrow \mathbf{L}$, where $\mathbf{L}$ is the matrix of output LLRs with dimensions specified below, we employ a batch normalized CNN with residual connections using a preactivation ResNet \cite{he2016identity}, as described in Table~\ref{tab:cnn_arch}. The applied ResNet blocks are described in detail in Fig.~\ref{fig:resnet}. The 2D convolutions operate over the first two dimensions, time and frequency. The architecture is fully convolutional and therefore the size of its output ($S\times F)$ is determined directly from the size of the input, and it can even vary for each TTI when using the same trained neural network. Namely, LLR estimates must be calculated for each symbol in the input and thus constant resolution ($S\times F)$ is maintained throughout the network without resorting to max-pooling or striding, which are often used in CNNs to reduce resolution. Also note that the input size can be larger than what was used in the training, but not smaller than the total receptive field because the zero-padding for the convolutions might produce unpredictable effects in the output.

   Instead of altering the resolution, we increase the amount of filters in the middle of the network and apply dilated convolutions to increase the receptive field as is often done, for example, in semantic segmentation \cite{Yu:2016:MCA}. Using dilation instead of striding enables the network to retain the detailed information about each input symbol while still allowing it to capture longer dependencies in time and frequency. The dilations proved to be especially important for the more shallow architectures. We also observed improved results when using depthwise separable convolutions  \cite{conf/cvpr/SifreM13,chollet2017} instead of normal convolutions. The main results in this paper are computed with a depth multiplier value of 2 for the depthwise separable convolutions, which simply means that the number of output channels in the depthwise convolution is doubled in order to increase the number of parameters and thereby improve the modeling capability of the network. However, it was observed that using the value of~1 works also well. Section~\ref{sec:ablation} contains a detailed ablation study related to these and various other architectural choices.

\begin{table}[!t]
	\renewcommand{\arraystretch}{1}
	
	\caption{The DeepRx CNN ResNet architecture. The ResNet block is described in Fig. \ref{fig:resnet}.} 
	\vspace{-0.1in}
	\label{tab:cnn_arch}
	\centering
	\footnotesize
	\begin{tabular}{l|l|l|l|l}
		\textbf{Layer} & \textbf{Type} & \textbf{Filter ($S$, $F$)}& \textbf{Dilation ($S$, $F$)} & \textbf{Output Shape}  \\ \hline\hline
		Input 1 $\mathbf{Y}\in\mathbb{C}$&  RX Data & & & ($S$, $F$, $N_r$) \\ \hline
		
		Input 2 $\mathbf{X}_p\in\mathbb{C}$&  TX Pilot & & & ($S$, $F$, $1$) \\ \hline
		Input 3 $\mathbf{H}_r\in\mathbb{C}$&  Raw channel estimate & & & ($S$, $F$, $N_r$) \\ \hline
		Input $\mathbf{Z}_c\in\mathbb{C}$&  Concatenate inputs 1-3 $\in\mathbb{C}$ & & & ($S$, $F$, $N_c$) \\ \hline
		Real input $\mathbf{Z}\in\mathbb{R}$&  Concatenate $\in\mathbb{R}$ & & & ($S$, $F$, $2N_c$) \\ \hline
		Conv In & 2D convolution & (3,3) & (1,1) & ($S$, $F$, 64) \\ \hline
		ResNet Block 1 &  Depthwise separable conv. & (3,3)& (1,1) & ($S$, $F$, 64) \\ \hline
		ResNet Block 2 &  Depthwise separable conv. & (3,3)& (1,1) & ($S$, $F$, 64) \\ \hline
		ResNet Block 3&  Depthwise separable conv. & (3,3)& (2,3) & ($S$, $F$, 128) \\ \hline
		ResNet Block 4 &  Depthwise separable conv. & (3,3)& (2,3) & ($S$, $F$, 128) \\ \hline
		ResNet Block 5 &  Depthwise separable conv. & (3,3)& (2,3) & ($S$, $F$, 256) \\ \hline
		ResNet Block 6 &  Depthwise separable conv. & (3,3)& (3,6) & ($S$, $F$, 256) \\ \hline
		ResNet Block 7 &  Depthwise separable conv. & (3,3)& (2,3) & ($S$, $F$, 256) \\ \hline
		ResNet Block 8 &  Depthwise separable conv. & (3,3)& (2,3) & ($S$, $F$, 128) \\ \hline
		ResNet Block 9 &  Depthwise separable conv. & (3,3)& (2,3) & ($S$, $F$, 128) \\ \hline
		ResNet Block 10 &  Depthwise separable conv. & (3,3)& (1,1) & ($S$, $F$, 64) \\ \hline
		ResNet Block 11 &  Depthwise separable conv. & (3,3)& (1,1) & ($S$, $F$, 64) \\ \hline
		Conv Out & 2D convolution & (1,1) & (1,1) & ($S$, $F$, $B$) \\ \hline
		LLR Output $\mathbf{L}$ & Output & & & ($S$, $F$, $B$) \\ \hline
	\end{tabular}
\end{table}

Finally, the prediction of bits is simply modeled as a binary classification problem. The final output of the DeepRx consists of the bit LLRs $\mathbf{L}\in\mathbb{R}^{S \times F \times B}$, where $B$ is the number of bits in the used constellation (e.g., 4 for 16-QAM). We use the binary sigmoid cross-entropy (CE) loss between each of the ground truth bits and the output $\mathbf{L}$ of the network,
\begin{align}
  {\mathrm{CE}}(\boldsymbol{\theta})\triangleq
  - \frac{1}{\#\mathcal DB}\sum_{(i,j)\in \mathcal D}\sum_{l=0}^{B-1}\left(b_{ijl} \log(\hat{b}_{ijl}) + (1 - b_{ijl}) \log(1 - \hat{b}_{ijl})\right)
\end{align}
where $\#\mathcal D$ is the number of resource elements carrying data and $\hat{b}_{ijl}$ is an estimate for the probability that the bit $b_{ijl}$ is one,
\begin{align}
  \hat{b}_{ijl} = \operatorname{sigmoid}\left(L_{ijl}\right) = \frac{1}{1 + e^{-L_{ijl}}} .
\end{align}

{\bf Remark:} Even though the actual bits are used as the ground truth of outputs in the cross-entropy loss, we consider the LLRs ($\mathbf{L}$) as the output of the inference network. The LLRs represent also the model's uncertainty about the bits and can be fed, for example, to an LDPC decoder, which then makes the decisions regarding the actual information bits.

\begin{figure}%
\vspace{-3.5mm}
	\centering
	\subfloat[The relationship between 16QAM and QPSK constellation points.]{{\includegraphics[width=0.56\linewidth]
			{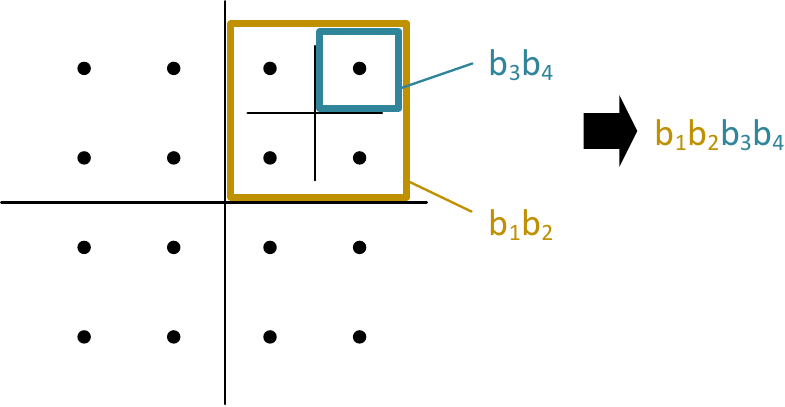} }}%
	\qquad
	\subfloat[Masking bits based on modulation.]{{\includegraphics[width=0.36\linewidth]
			{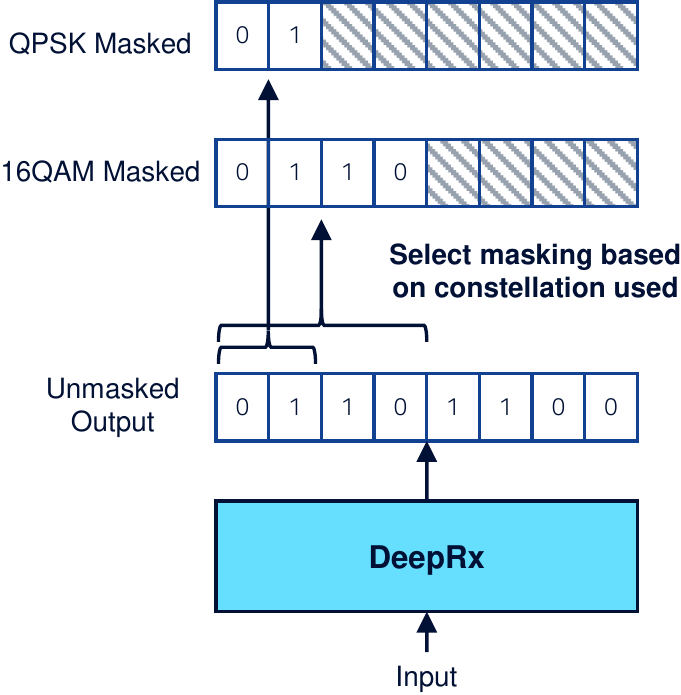} }}%
                    \caption{(a) The hierarchical relationship between QPSK and 16-QAM: In QPSK, the two bits b$_1$ and b$_2$, are defined by the complex quadrant of the symbol. In 16-QAM, this also holds for the first two bits (orange box), while the last two bits, b$_3$ and b$_4$, correspond to the finer location within the quadrant (blue box). (b) Bit masking for supporting multiple QAM modulations in DeepRx.}%
	\label{fig:constellations}%
\end{figure}

The output of DeepRx has been designed so that a single network can be trained to support multiple quadrature amplitude modulation (QAM) schemes. As shown in Fig.~\ref{fig:constellations}a, the QAM schemes in 5G (QPSK, 16-QAM, 64-QAM, 256-QAM) are related to each other hierarchically so that it is possible to map four constellation points in the higher-order modulation to a single constellation point in the lower-order modulation. We take advantage of this relation and define the output of the model in such a way that same output bits/LLRs correspond to the same part of the constellation space, regardless of the modulation used. This means that, for an individual symbol, the first two outputted bits describe the complex quadrant of the symbol (1st-order point), the next two bits define the quadrant around the 1st-order point (giving the 2nd-order point), etc. In practice, this is achieved by setting the number of outputs according to the highest supported modulation order, in this case choosing 8 outputs to support up to 256-QAM modulation. During training and inference, the outputs are then masked, as depicted in Fig. \ref{fig:constellations}b, such that only the bit positions actually used in the modulation scheme are considered. During training, the masked-out bits do not affect the computed loss for that symbol, while during inference the masked-out bits are simply omitted. With this approach, a single network can learn to support all modulation orders. 

\section{Generation of Training and Validation Data}
\label{sec:training_data}

\begin{table}
  \setlength{\tabcolsep}{3pt}
    \renewcommand{\arraystretch}{1.2}
    \footnotesize
    \centering
    \caption{Simulation parameters for training and validation.}
    \begin{tabular}{l|p{3.2cm}|p{3.2cm}|l}
    \hline
    \multicolumn{1}{c|}{\textbf{Parameter}} & \textbf{Training} & \textbf{Validation} & \textbf{Randomization}\\
		\hline\hline
		Carrier frequency & \multicolumn{2}{c|}{4 GHz} & None\\
    \hline
		Channel model & CDL-B, CDL-C, CDL-D, TDL-B, TDL-C, TDL-D & CDL-A, CDL-E, TDL-A, TDL-E & Uniform\\
		\hline
		RMS delay spread & \multicolumn{2}{c|}{10 ns -- 300 ns} & Uniform\\
		\hline
		Maximum Doppler shift & \multicolumn{2}{c|}{0 Hz -- 500 Hz} & Uniform\\
		\hline
		SNR & \multicolumn{2}{c|}{$-4$ dB -- $32$ dB} & Uniform\\
		\hline
		SIR & \multicolumn{2}{c|}{$0$ dB -- $36$ dB} & Uniform\\
		\hline
		Number of PRBs & \multicolumn{2}{c|}{26 (312 subcarriers)} & None\\
		\hline
		Subcarrier spacing & \multicolumn{2}{c|}{15 kHz} & None\\
		\hline
		OFDM symbol duration & \multicolumn{2}{c|}{71 $\mu$s} & None\\
		\hline
		TTI length & \multicolumn{2}{c|}{14 OFDM symbols / 1 ms} & None\\
		\hline
		Modulation scheme & \multicolumn{2}{c|}{16-QAM} & None\\
		\hline
		Code rate & \multicolumn{2}{c|}{$\frac{658}{1024}$} & None\\
		\hline
		Number of RX/TX antennas & \multicolumn{2}{c|}{2/1} & None\\
		\hline
		DMRS configuration & \multicolumn{2}{c|}{Four options, see Fig.~\ref{fig:pilotconfig}} & Uniform\\
		\hline
    \end{tabular}
    \label{table:param}
  \end{table}

The training and validation data is generated with the link-level simulator implemented with Matlab's 5G Toolbox, which is modeling a 5G PUSCH link as described in Section \ref{sec:simulator}. The parameter values used in the simulations are listed in Table~\ref{table:param}. Each individual data set contains 500~000~TTIs, of which 60\% is used for training, and a subset of the remaining 40\% is used for validation. As already mentioned, the HARQ procedure is purposefully omitted from the simulations since it has implications only for the processing occurring after calculating the LLRs. This renders the proposed DeepRx architecture transparent to the HARQ process.

The randomization of the parameters is done for each frame (period of 10 TTIs), using the ranges and distributions indicated in Table~\ref{table:param}. Also the signal-to-noise ratio (SNR) and signal-to-interference ratio (SIR) are randomized, although it should be noted that interference is present only in some of the results. Moreover, we wish to emphasize that drawing the logarithmic SNR from a uniform distribution is a conscious choice to make the training more efficient for the full considered SNR range. It is well known that in reality SNR is more likely to follow a log-normal distribution.

\begin{figure*}[t!]
	\centering
	\includegraphics[width=0.8\textwidth,trim={2.5cm 11.9cm 2.5cm 12cm},clip]{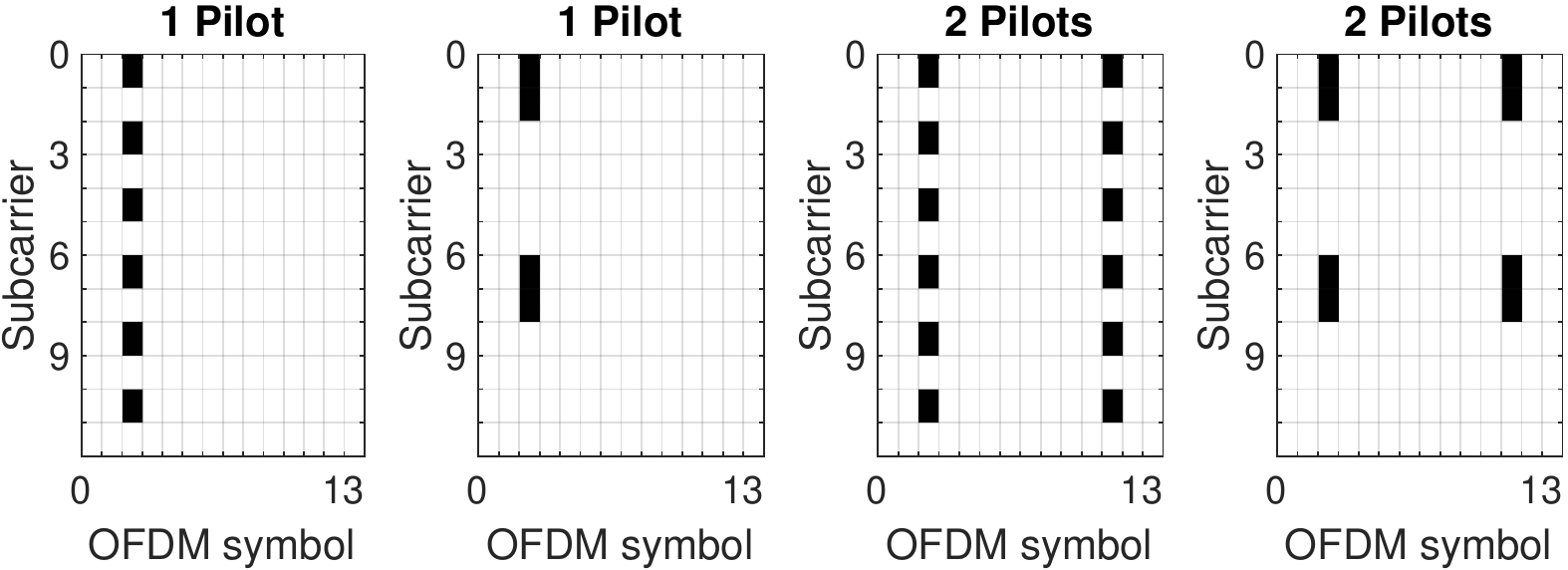}
	\caption{The considered DMRS/pilot configurations, illustrated for one PRB over the duration of a TTI. Note that in the forthcoming results the pilot configurations are only differentiated in terms of how many OFDM symbols they utilize.}
	\label{fig:pilotconfig}
\end{figure*}

As for the DMRS configuration, four different options are included in the data, each of them illustrated in Fig.~\ref{fig:pilotconfig} for a single PRB. For each frame of 10 TTIs, one of these configurations is chosen randomly. Note that in the validation results the four pilot configurations are only differentiated in terms of how many OFDM symbols include pilots (i.e., either one or two). We refer to these cases as ``one pilot`` or ``two pilots'' in the following. The reason for this is that we have not observed the frequency domain positioning of DMRS symbols to have any significant impact on the performance, and differentiating between them in the figures would only clutter the results without bringing any new insights. It should also be noted that only the DMRS configurations with two pilots are feasible for traditional LMMSE receivers, given the rather large Doppler range. The configurations with one pilot are included in the study only in order to evaluate the performance boundaries of DeepRx.

The simulated data sets are generated using ten different 3GPP channel models, whose detailed descriptions can be found in \cite{NR_38901}. The models ending with letters \emph{A}, \emph{B}, and \emph{C} represent NLOS, while those ending with \emph{D} and \emph{E} are constructed for LOS. It is also important to note that the models \emph{CDL-A}, \emph{CDL-E}, \emph{TDL-A}, and, \emph{TDL-E} are used only for validation, while the training is performed with the remaining six channel models listed in Table~\ref{table:param} (hence the 60\%/40\% split between training and validation data). This ensures that the deep neural network cannot achieve its high performance by simply learning the characteristics of the individual channel models.

In addition, we also generate a smaller training data set using a fully synthetic channel model, where the channel for an individual OFDM symbol is a 7-tap Rayleigh fading channel. After each OFDM symbol, the channel is randomly changed such that 90\% of the variance of the new channel consists of the previous channel realization, while 10\% of the variance is stemming from a new randomly generated 7-tap channel. This ensures a certain level of channel correlation between consecutive OFDM symbols. The purpose of this artificial channel model is not to represent a realistic propagation channel, but simply capture some fundamental properties of a wireless channel while still being different from the realistic 3GPP channel models. This allows for some insightful experiments as reported in more detail below in Section~\ref{sec:expl-perf}. However, it should be emphasized that in the forthcoming results the 3GPP channels are used for both training and validation unless mentioned otherwise.

\section{Results}
\label{sec:validation_res}

In this section, we compare the proposed CNN-based receiver, DeepRx, to two traditional LMMSE (Section~\ref{sec:simulator}) receivers:
\begin{itemize}
\item One that performs LS channel estimation and interpolates the channel estimate over the data symbols and subcarriers;
\item One that obtains the full channel information as a~priori knowledge.
\end{itemize}
The former represents a realistic benchmark and is therefore referred to as a practical LMMSE receiver, while the latter one estimates the upper bound of the achievable performance with LMMSE equalization. Since we observed that approximating the LLRs with \eqref{eq:llr_appr} has only a negligible impact on the final receiver performance, the approximated demapping rule is used in both the benchmark receivers, including the one with full channel knowledge. It should also be emphasized that the LMMSE receiver with full channel knowledge is resorting to otherwise practical OFDM receiver processing, meaning that it is not always able to achieve perfect equalization. This is primarily due to inter-carrier-interference caused by Doppler spread and the underlying channel changing also within each resource element.

We use bit error rates (BER) as the main performance criteria throughout this section. In particular, we consider two types of BERs: 1) a ``raw'' BER based on the hard decision (bit is 1 if LLR $<$ 0, or 0 otherwise) referred to as the uncoded BER, and 2) the coded BER obtained by feeding the LLRs through a 5G-compliant LDPC decoder (preceded by a rate dematcher) and comparing the decoded bits to the original bit sequence. Investigating the coded BER reveals whether the LLRs provided by the DeepRx sufficiently capture the uncertainty of the detected bits for the purposes of LDPC decoding. The coded BER can also be seen as a proxy to the accuracy of the LLRs themselves, as direct evaluation of LLRs is difficult due to lack of ground truth (there is no explicit formula for the ideal LLRs under the utilized channel models).

The model is trained using the simulated link-level data described in Section~\ref{sec:training_data}. The optimization is carried out using the LAMB optimizer \cite{You2020Large}, starting from a random initialization with the main learning rate set to  $10^{-2}$. We also apply a small weight decay with the scaling factor of $10^{-4}$ that prevents the weight magnitudes from growing during a long training run. LAMB allows for scaling the training to larger batch sizes (e.g., 80 TTIs, each TTI having $312 \times 14 \times 8$ bits, altogether $\sim$2,8M bits per batch, using four 2080Ti GPUs in parallel), while for smaller batch sizes (e.g., 20 TTIs), AdamW \cite{loshchilov2017decoupled} might also suffice. For large batch sizes, we also use a linear learning rate warmup period from zero to main learning rate, with the duration of 800 iterations. In addition, the learning rate is decayed linearly to zero after reaching 30\% of the total iterations. In total, we usually run 30k iterations with a batch size of 80 TTIs, and we have not observed significant improvement with longer runs. Overfitting was not observed with our models using the datasets described above, meaning that DeepRx was not observed, for instance, to memorize the training samples and thereby compromise its performance under validation.

\subsection{Primary Validation Results}

\begin{figure}%
\vspace{-3.5mm}
	\centering
	\subfloat[Uncoded BER]{{\includegraphics[width=0.46\linewidth]
			{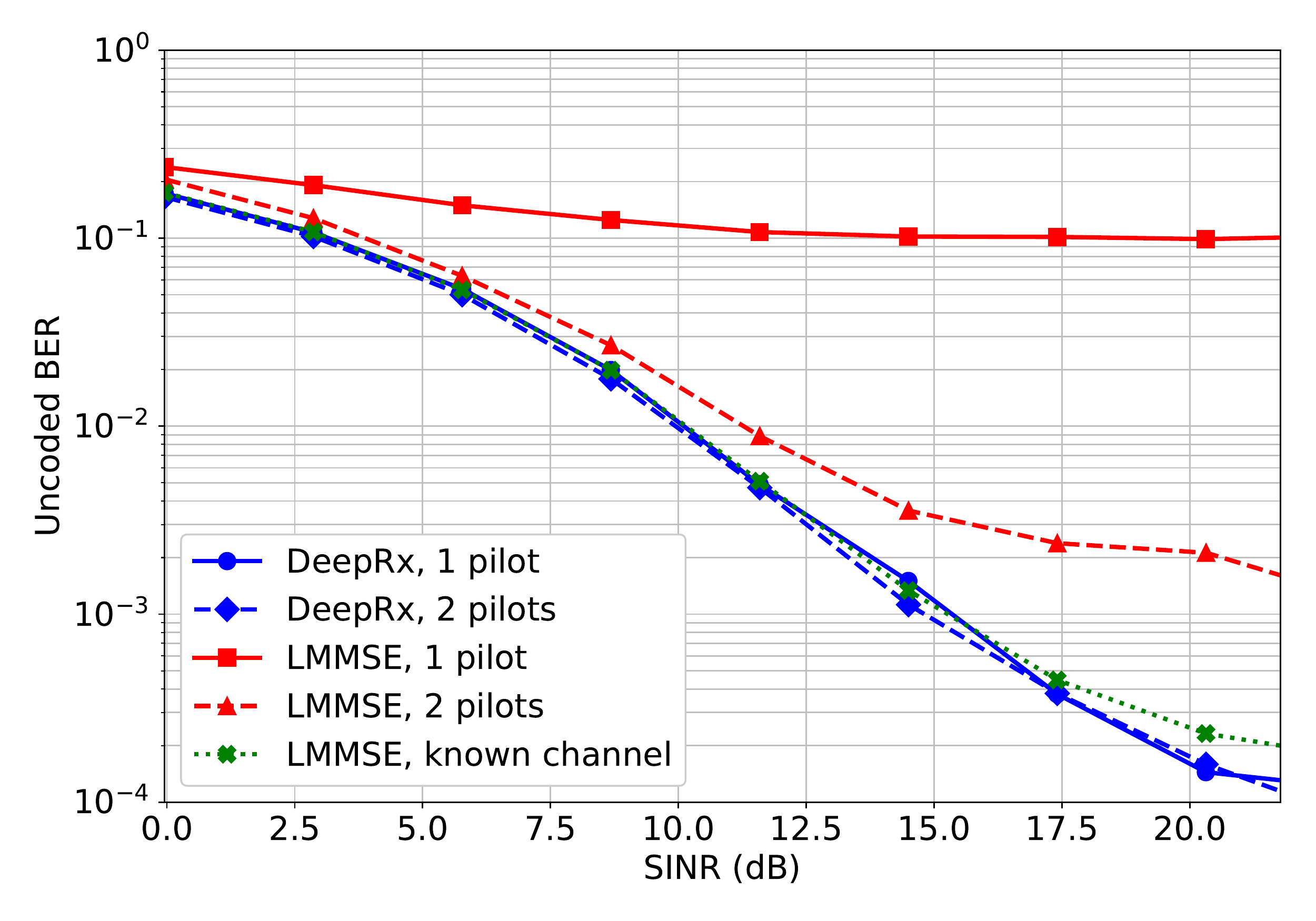} }}%
	\qquad
	\subfloat[Coded BER]{{\includegraphics[width=0.46\linewidth]
			{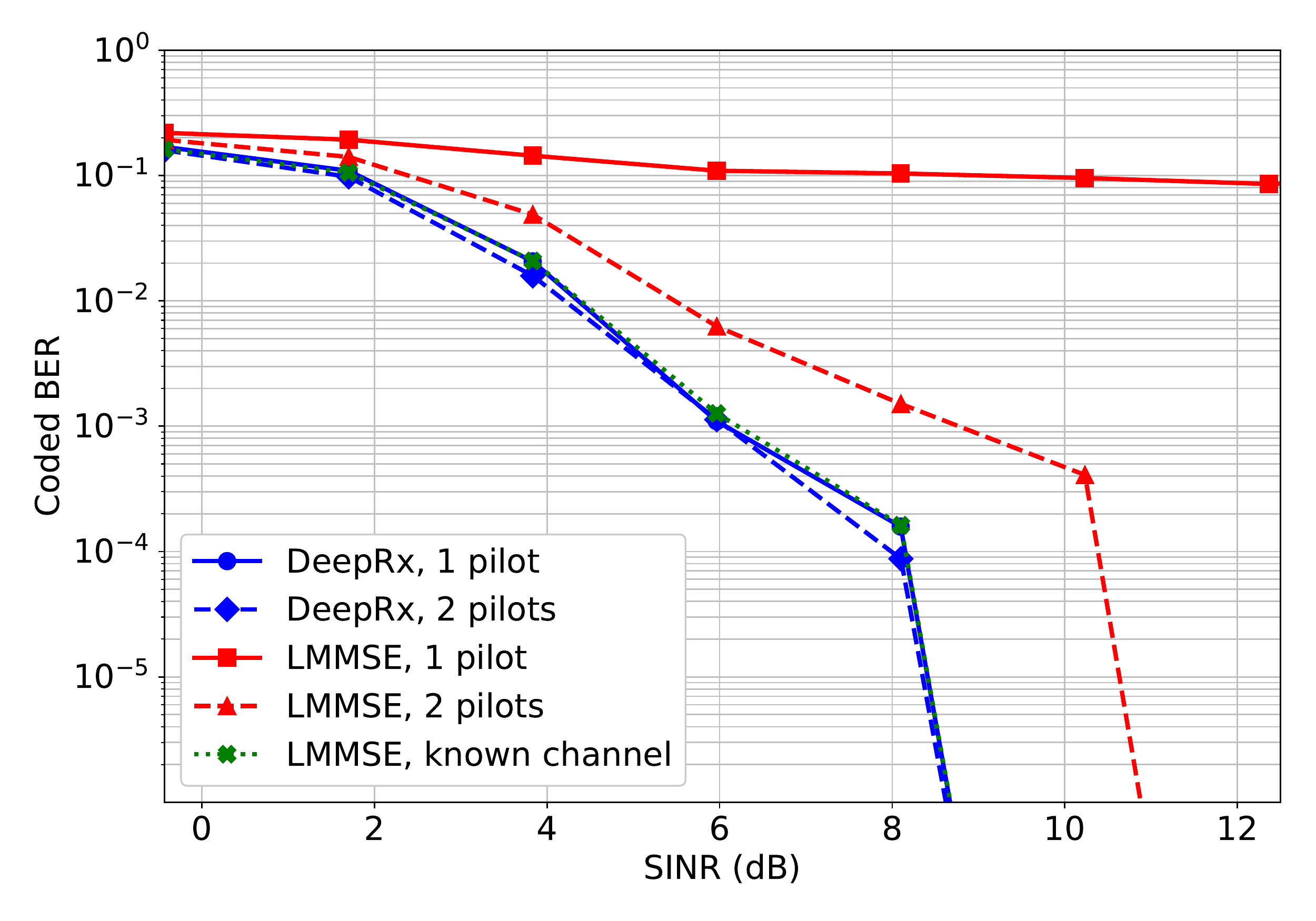} }}%
	\caption{(a) Uncoded BER and (b) coded BER performance of the DeepRx compared to the reference LMMSE receivers, without inter-cell-interference.}%
	\label{fig:noint_allpilots}%
\end{figure}

First, we consider performance of DeepRx without inter-cell-interference. From Fig.~\ref{fig:noint_allpilots}a, which shows the uncoded BER, it can be seen that DeepRx clearly outperforms the practical LMMSE receiver performing LS channel estimation. In fact, the CNN-based DeepRx can essentially match the performance of the LMMSE receiver with \emph{full channel knowledge}, even when it has just one pilot symbol in time at its disposal. Due to the rather wide Doppler shift range within the data, the practical LMMSE receiver performs very poorly with just one pilot since it requires two pilots to operate reliably under such conditions. This is in stark contrast to DeepRx, for which, at higher SINRs, just one pilot is enough to outperform by a factor of 10 the practical LMMSE receiver having two pilots at its disposal. The reasons behind such exceptionally high performance are investigated further in Section \ref{sec:expl-perf}.

In addition, the coded BERs (Fig.~\ref{fig:noint_allpilots}b) indicate that the LLRs calculated by DeepRx are of sufficient quality for the LDPC decoder as the coded BER matches that of the LMMSE receiver with full channel knowledge. The gain over the practical LMMSE receiver with two pilots is roughly 2~dB, whereas the practical LMMSE receiver does not even reach the waterfall region of the code with one pilot. The performance of DeepRx suffers only a marginal reduction when the number of pilot symbols is reduced from two to one.

\begin{figure}%
\vspace{-3.5mm}
	\centering
	\subfloat[Uncoded BER]{{\includegraphics[width=0.46\linewidth]
			{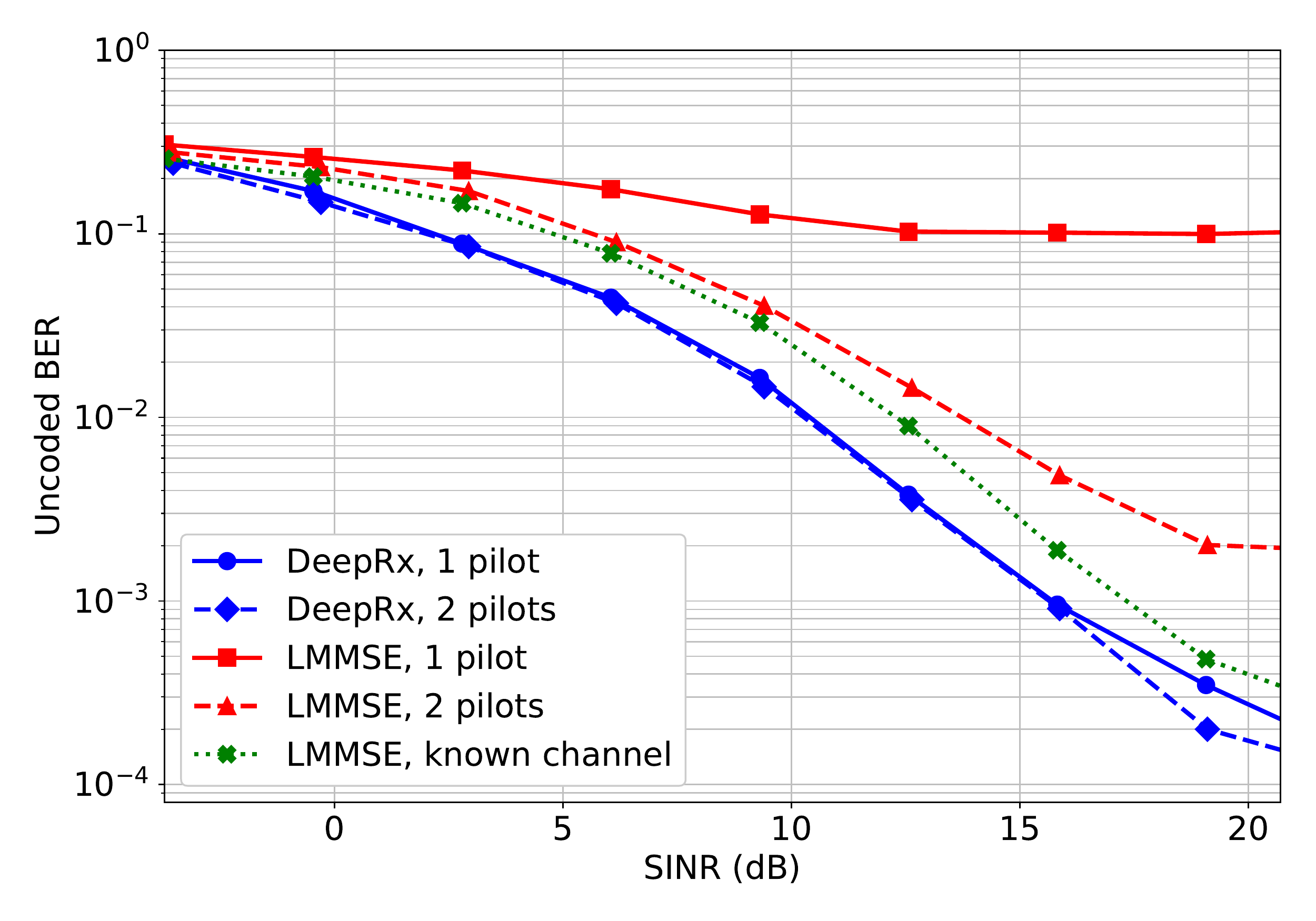} }}%
	\qquad
	\subfloat[Coded BER]{{\includegraphics[width=0.46\linewidth]
			{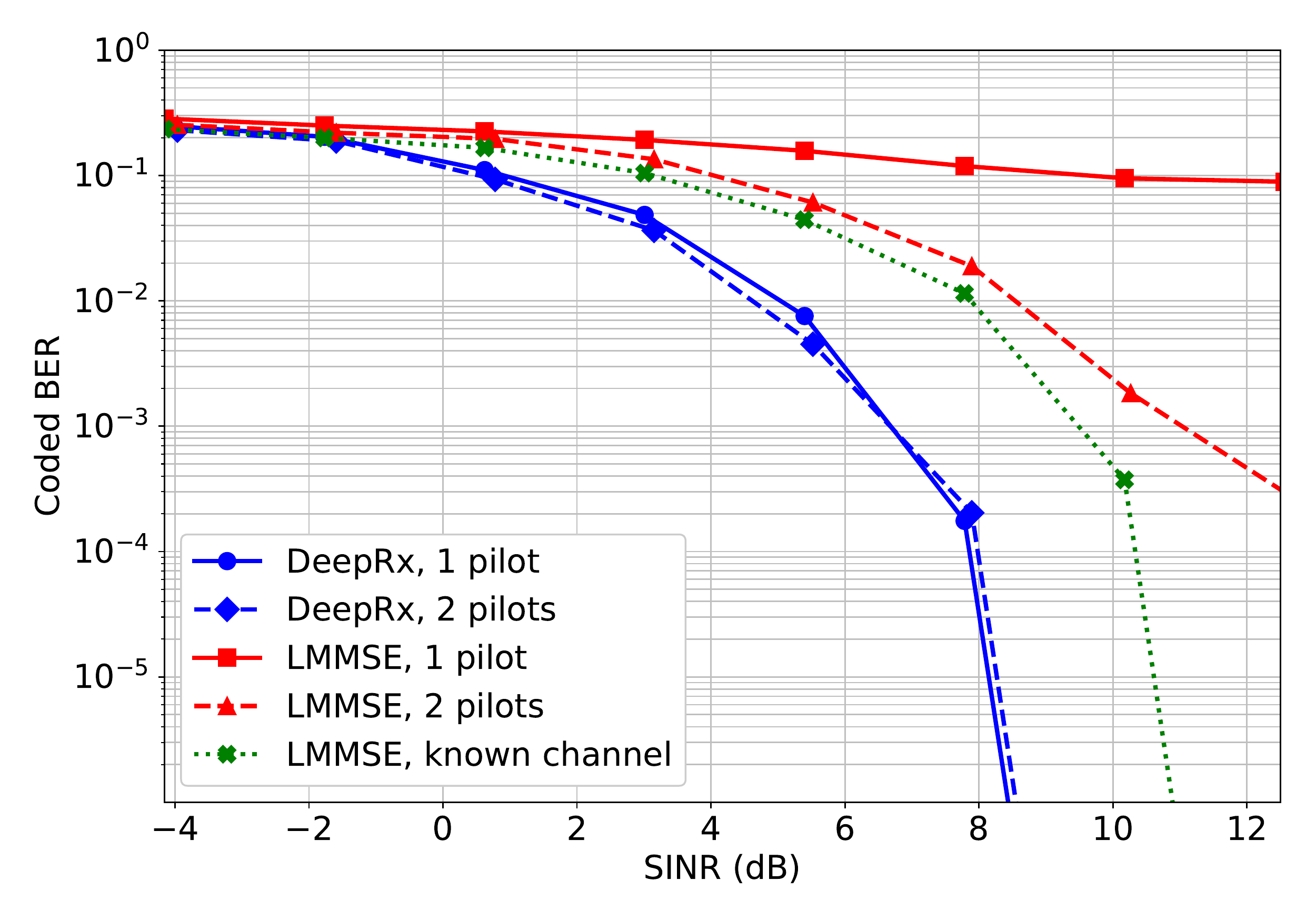} }}%
	\caption{(a) Uncoded BER and (b) coded BER performance of the DeepRx compared to the reference LMMSE receivers, with inter-cell-interference.}%
	\label{fig:int_allpilots}%
\end{figure}

Next, we consider a case with some inter-cell-interference in the received uplink signals. The interfering signal has a similar waveform as the signal of interest, but it has a random time offset and a different channel realization. The power of the interference is on average 4~dB lower than the noise power. The results are shown in Fig.~\ref{fig:int_allpilots}. It can be observed that all the reference LMMSE receivers suffer from reduced performance as they do not have any capabilities for minimizing the effects of the interference. On the other hand, since DeepRx is trained with data that includes similar interference levels, it implicitly learns to manage the interference and outperforms even the LMMSE receiver with full channel knowledge. However, we wish to emphasize that this interference-related performance gap could likely be reduced if the LMMSE receivers were utilizing, for example, some type of interference rejection combining (IRC) \cite{Leost12a}.

Under inter-cell-interference, the performance gain of DeepRx compared to the reference receivers is even higher when coded BER is considered (Fig.~\ref{fig:int_allpilots}b). This likely stems from the fact that the traditional demapper assumes Gaussian-distributed white noise in the calculation of LLRs, while the rather strong interference signal invalidates this assumption. DeepRx does not resort to such assumptions as it implicitly learns the proper receiver procedure for the given noise-plus-interference distributions, based on the training data it observes. Consequently, it outperforms even the benchmark with full channel knowledge by 2~dB, while the practical LMMSE receiver fails to even enter the waterfall region of the code within the considered SINR range.

\subsection{Exploring the Reasons Behind DeepRx's Performance}
\label{sec:expl-perf}

The exceptionally high performance of DeepRx obviously raises questions as to what the CNN actually learns to do. In this section we will discuss some of our experimental findings that provide some insight into this.

\subsubsection*{Temporal Tracking of the Channel}

\begin{figure*}[t!]
	
	\centering
	\includegraphics[width=0.56\linewidth]{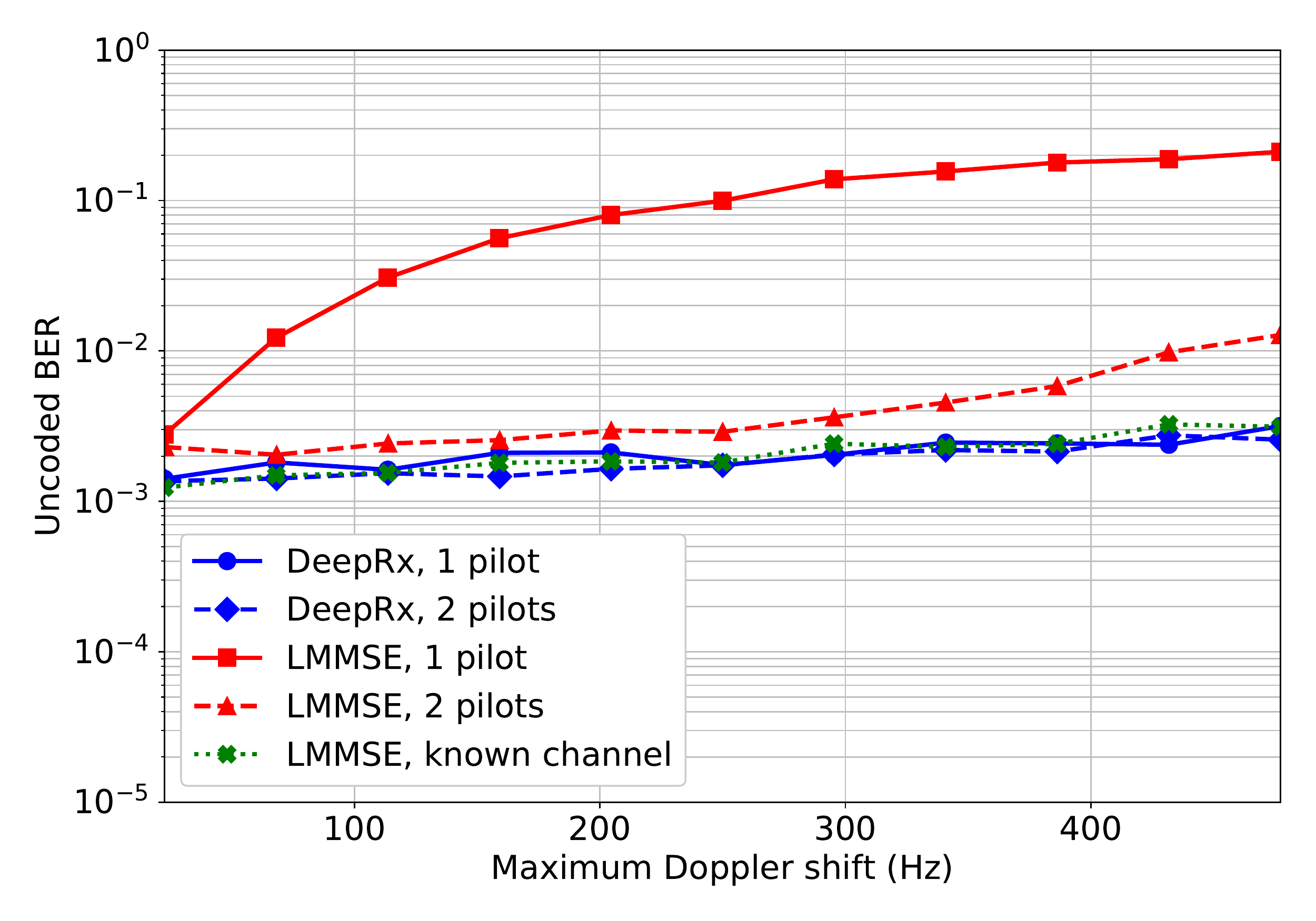}
	\caption{Uncoded BER with respect to the maximum Doppler shift of the channel, calculated for SINRs sampled uniformly between 10--20 dB and without inter-cell-interference.}%
	\label{fig:doppler}
\end{figure*}

Figs.~\ref{fig:noint_allpilots} and~\ref{fig:int_allpilots} showed very high performance with just one pilot symbol, positioned in the first half of the TTI. This suggests that the CNN-based DeepRx is capable of exceptionally accurate temporal tracking of the channel. To investigate this aspect in more detail, Fig.~\ref{fig:doppler} shows the uncoded BER with respect to the maximum Doppler shift of the channel, which is inversely proportional to the coherence time of the channel. As can be expected, the practical LMMSE receiver with one pilot suffers from a rapid degradation of BER when the Doppler shift increases, as it must assume that the channel remains constant throughout the TTI. With two pilots, even the practical LMMSE receiver is capable of tracking the channel rather well up to Doppler shifts of 400~Hz, although DeepRx outperforms it throughout the considered Doppler shift range. In fact, even with the highest considered Doppler shift of 500~Hz, DeepRx using just one pilot symbol can match the BER of the LMMSE receiver with full channel knowledge. This corresponds to a UE velocity of roughly 135~km/h, representing already a case of rather severe mobility. Note that the slight increase of BER for DeepRx and LMMSE receiver with full channel knowledge under the higher Doppler shifts is likely due to the frequency-domain spreading of the subcarriers, which reduces their orthogonality.

\begin{figure}%
\vspace{-3.5mm}
	\centering
	\subfloat[Uncoded BER]{{\includegraphics[width=0.46\linewidth]
			{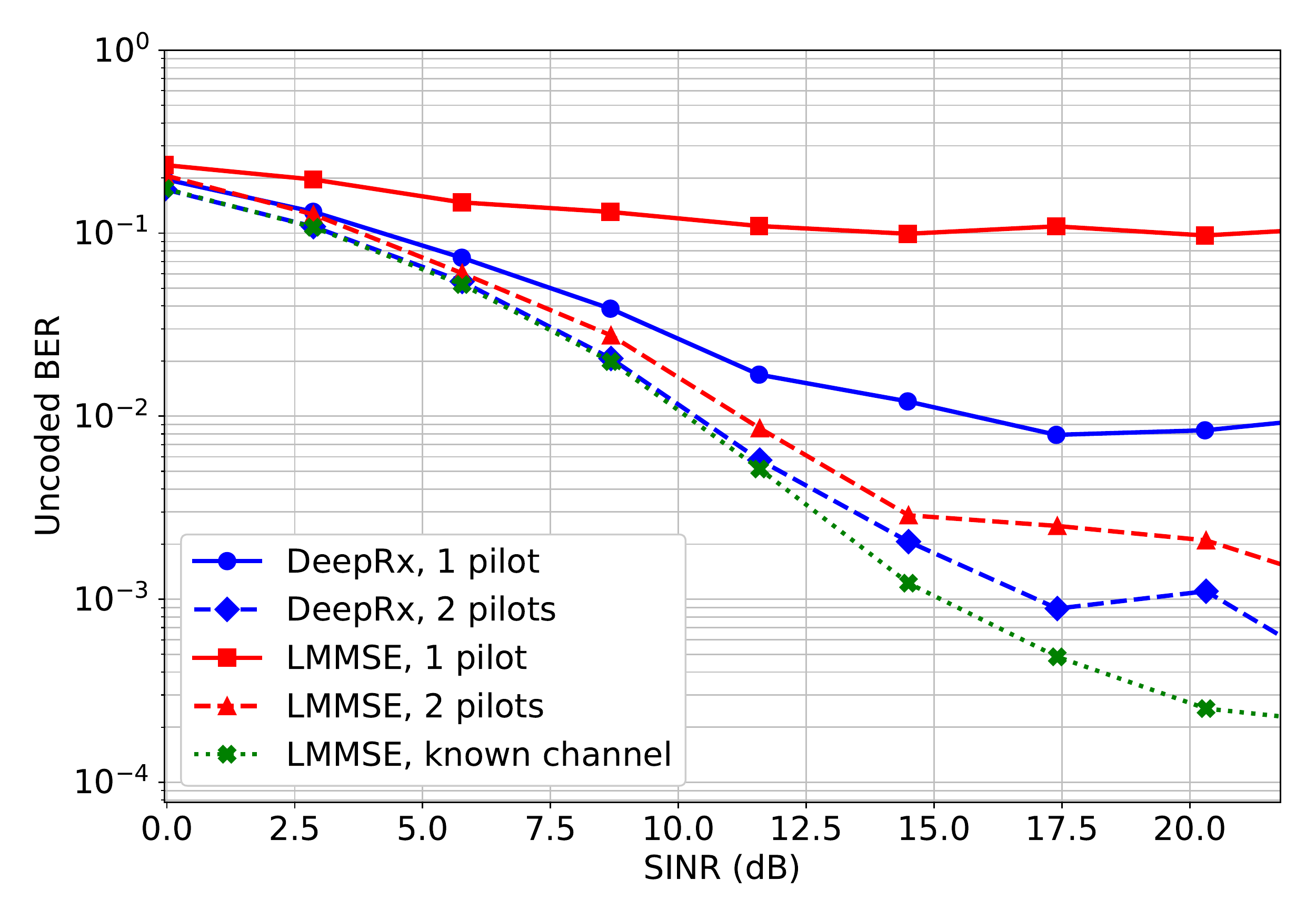} }}%
	\qquad
	\subfloat[Coded BER]{{\includegraphics[width=0.46\linewidth]
			{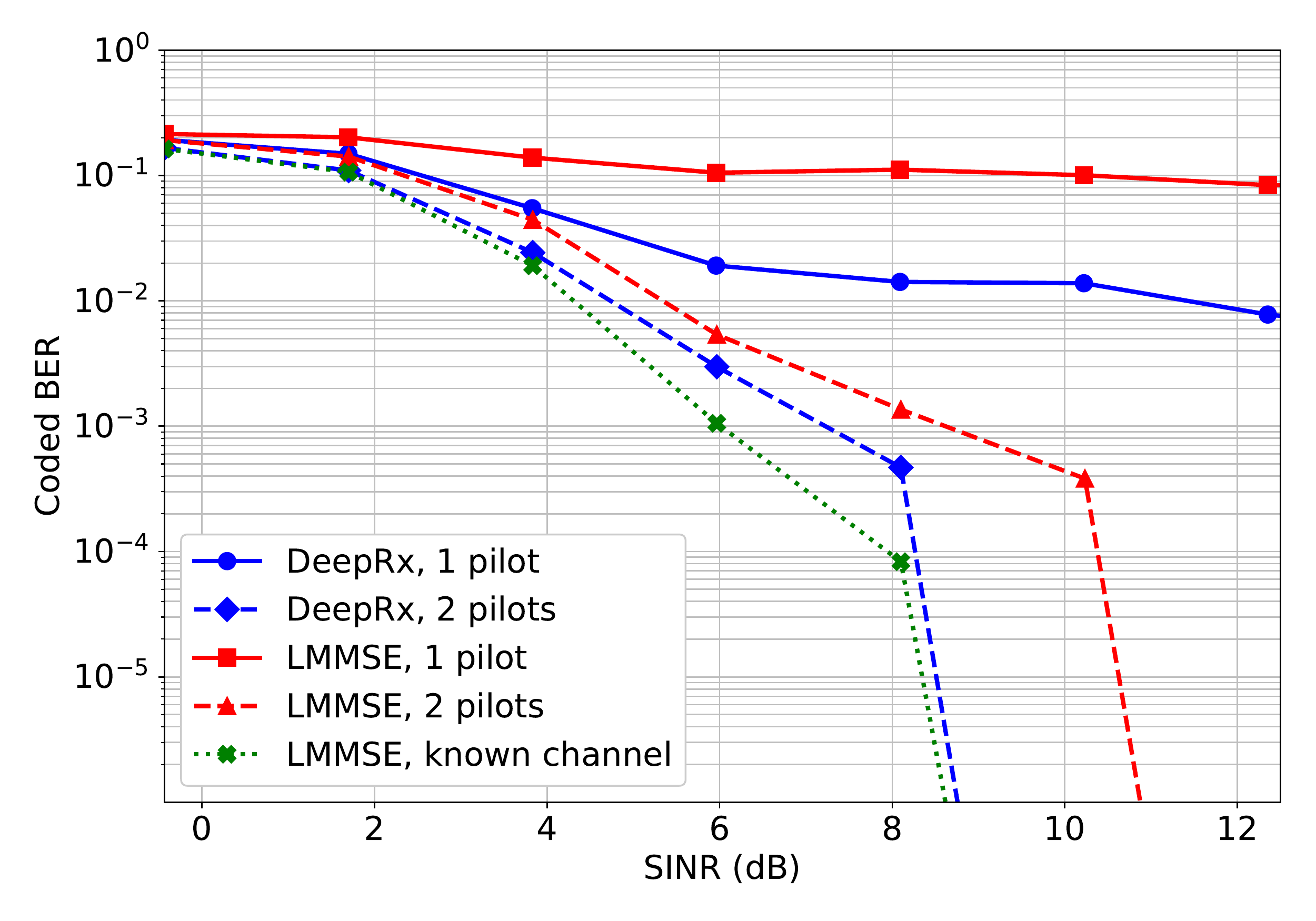} }}%
	
	\caption{(a) Uncoded BER and (b) coded BER performance of the DeepRx trained with a synthetic channel model and validated using the 3GPP channel models, compared to the reference LMMSE receivers, without inter-cell-interference.}%
	\label{fig:trainsynth_valid3gpp}%
\end{figure}

\subsubsection*{DeepRx Does Not Cheat by Learning Channel Models}

Even though the channel models used in training are different from those used for validation, which prevents DeepRx from taking unfair advantage of the properties of individual channel models, it is still a possibility that the channel models share some modeling deficiencies which it can exploit. In order to ensure that no such exploitation occurs, we also trained DeepRx using the synthetic Rayleigh channel model described in Section~\ref{sec:training_data}, which is completely independent from the 3GPP channel models. Figure~\ref{fig:trainsynth_valid3gpp} shows the performance of such model validated with the 3GPP channel models listed in Table~\ref{table:param}. Although the performance of the newly trained DeepRx is somewhat worse than when trained with the realistic 3GPP channel models, it still clearly outperforms the practical LMMSE receiver. For instance, with one pilot symbol, its BER performance is still an order of magnitude better than that of the practical LMMSE receiver. These results indicate that DeepRx indeed learns a generic solution for bit detection as it can be successfully applied to such different channel models. It is also likely that the performance deficit of the model trained with synthetic data is simply a result of the artificial behavior of the synthetic channel model, which prevents DeepRx from learning a fully accurate interpolation rule for the channel estimate\footnote{We wish to emphasize that we cannot claim that a neural network would carry out channel estimation or equalization in the traditional sense. Determining exactly what happens inside a trained neural network is an open question. When referring to DeepRx, we simply use these terms to reflect on the internal processing of pilots and RX data which is necessary for performing bit detection.}.

\begin{figure*}[t!]
	\centering
	\includegraphics[width=0.8\textwidth]{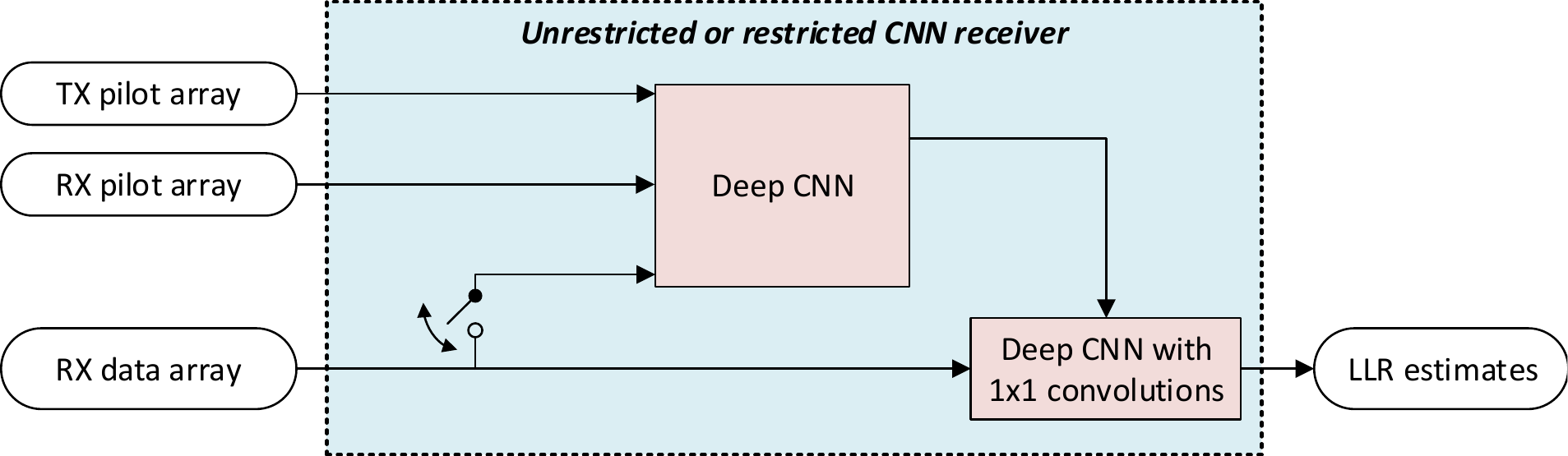}
	\caption{Two alternative CNN architectures for investigating the behavior of DeepRx.} 
	\label{fig:chan_est_diag}
\end{figure*}

\subsubsection*{Blind Utilization of the Unknown Data During the Detection Process}

To gain more insight into the reasons behind DeepRx's performance, let us train a more restricted CNN architecture. Fig.~\ref{fig:chan_est_diag} shows two alternatives where the RX data is routed differently through CNN. The two alternatives differ in whether or not the data-carrying RX subcarriers are fed to a deep CNN whose receptive field covers several subcarriers and symbols in time. This is depicted with a simple switch in Fig.~\ref{fig:chan_est_diag}, where a closed switch indicates that the deep CNN has full view of also the RX data array, in addition to the pilots. If the switch is open, on the other hand, the deep CNN has only access to the received and transmitted pilots, based on which it can perform the channel estimation. The equalization and demapping is carried out by a head of 3 layers of 1x1 convolutions (32 channels each), whose weights are also learned from the data. Such 1x1 convolutions facilitate equalization and demapping only in a symbol-by-symbol manner, i.e., the restricted CNN cannot utilize any spatial or temporal correlations in the RX data to improve the bit estimates. Furthermore, we wish to point out that the CNN architecture of Fig.~\ref{fig:chan_est_diag} with the switch closed is essentially identical in performance and architecture to the primary DeepRx architecture presented in Section~\ref{sec:nn_rx}, except for the additional 1x1 convolutional layers, which were added to ensure that it can be directly compared with the restricted CNN architecture. Therefore, for clarity, we shall refer to the unrestricted case as DeepRx in the forthcoming discussion. Moreover, we wish to point out that the restricted CNN architecture of Fig.~\ref{fig:chan_est_diag} is somewhat representative of the existing deep learning based channel estimators \cite{neumann18,he18}, and therefore the corresponding performance comparisons will provide also some insight into the benefits of DeepRx when compared against these prior deep learning-aided receivers.

\begin{figure}%
\vspace{-3.5mm}
	\centering
	\subfloat[Uncoded BER]{{\includegraphics[width=0.46\linewidth]
			{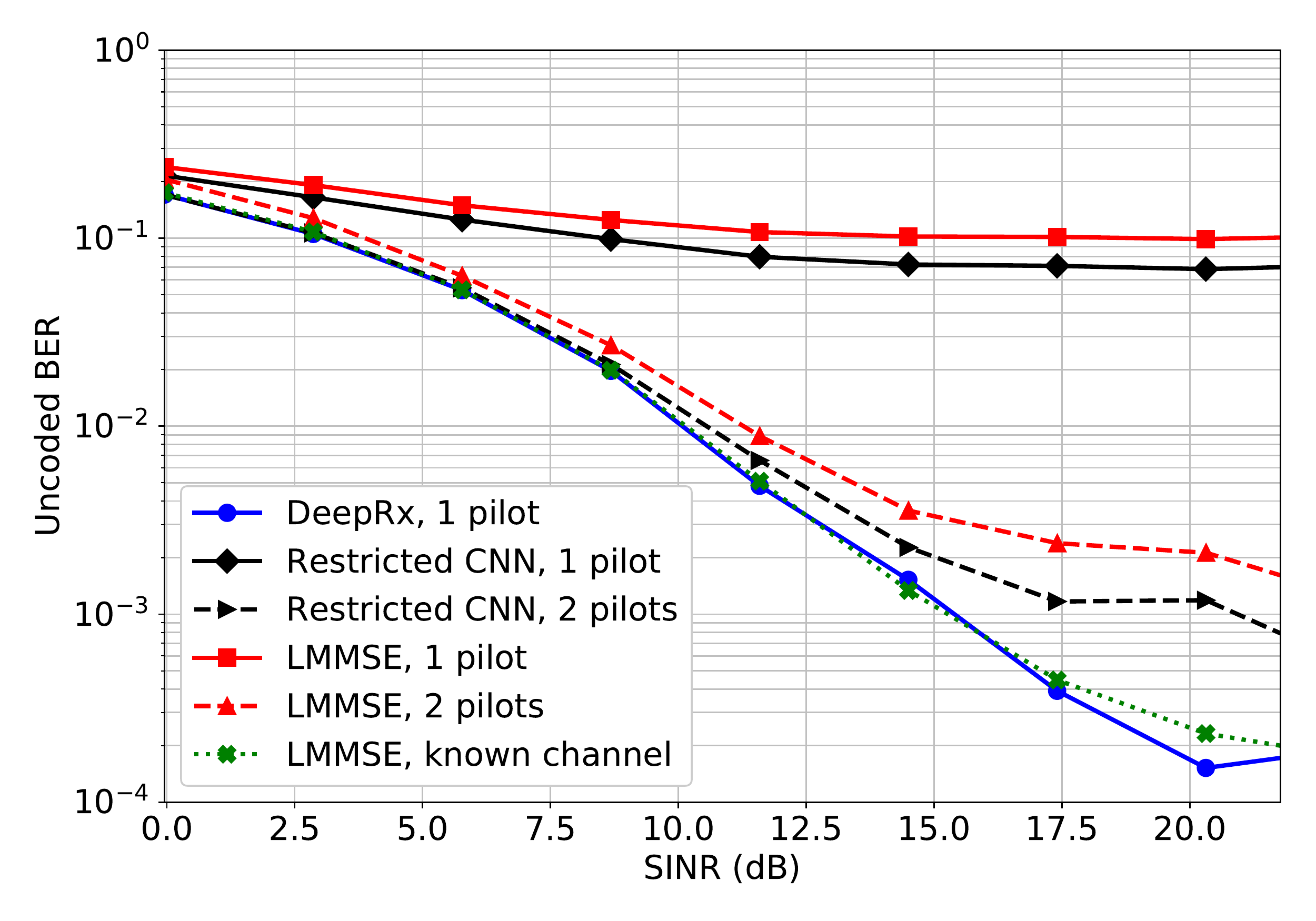} }}%
	\qquad
	\subfloat[Coded BER]{{\includegraphics[width=0.46\linewidth]
			{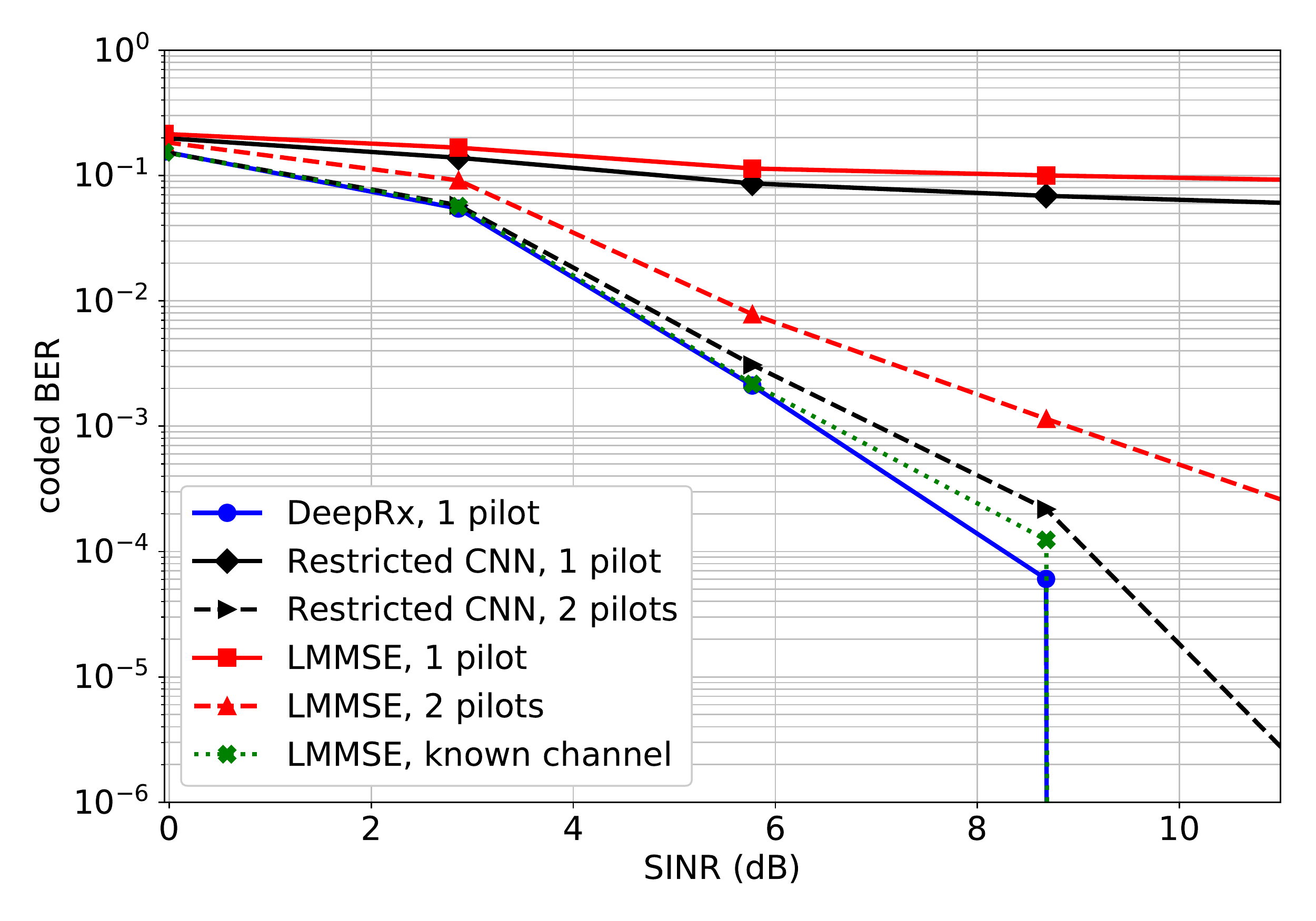} }}%
	
	\caption{(a) Uncoded BER and (b) coded BER performance of the restricted CNN Receiver which can only process one data symbol at a time after pilot-based channel estimation, compared to the the unrestricted DeepRx and reference LMMSE receivers, without inter-cell-interference.}%
	\label{fig:restricted_nn}%
\end{figure}

The BER results corresponding to the two architectures of Fig.~\ref{fig:chan_est_diag} are shown in Fig.~\ref{fig:restricted_nn}. For simplicity, the BER of DeepRx is only shown for the case of one pilot to ensure the readability of the figure. Altogether, it can be observed that the performance gain of the restricted CNN receiver over the practical LMMSE receiver is rather marginal. With one pilot, the BER of the restricted CNN receiver remains very high, whereas the unrestricted DeepRx is on par with the LMMSE having full channel knowledge, as observed already earlier. When there are two pilot symbols, the restricted CNN receiver fares better but is still clearly outperformed by DeepRx. These observations indicate that a crucial aspect for the high performance of DeepRx is to ensure that the CNN has full access to the data subcarriers. In other words, one should not impose too many restrictions on how the deep CNN processes the RX signal to obtain the LLR estimates, such as limiting the deep learning processing to only channel estimation. Namely, it is to be expected that considering all the tasks jointly results in higher performance than learning and performing them separately. This way DeepRx can freely learn an extremely accurate reception procedure which might differ from the processing flow of a traditional receiver. Finally, we note that we experimented also with larger and deeper 1x1 convolutional heads, and the results were similar, so the aforementioned performance differences cannot be attributed to a smaller network handling the data symbols.

\begin{figure*}[t!]
	\centering
	\includegraphics[width=0.35\textwidth]{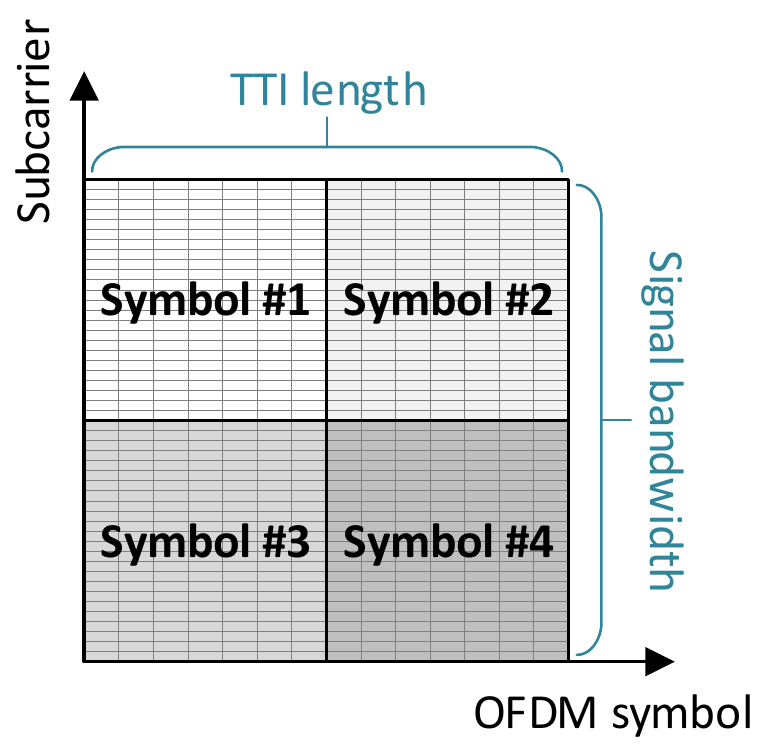}
	\caption{Artificial transmit symbol distribution used as validation data for probing the behavior of DeepRx.} 
	\label{fig:corner_illustration}
\end{figure*}

\begin{figure}%
\vspace{-3.5mm}
	\centering
	\subfloat[Regular QPSK data]{{\includegraphics[width=0.46\linewidth]
			{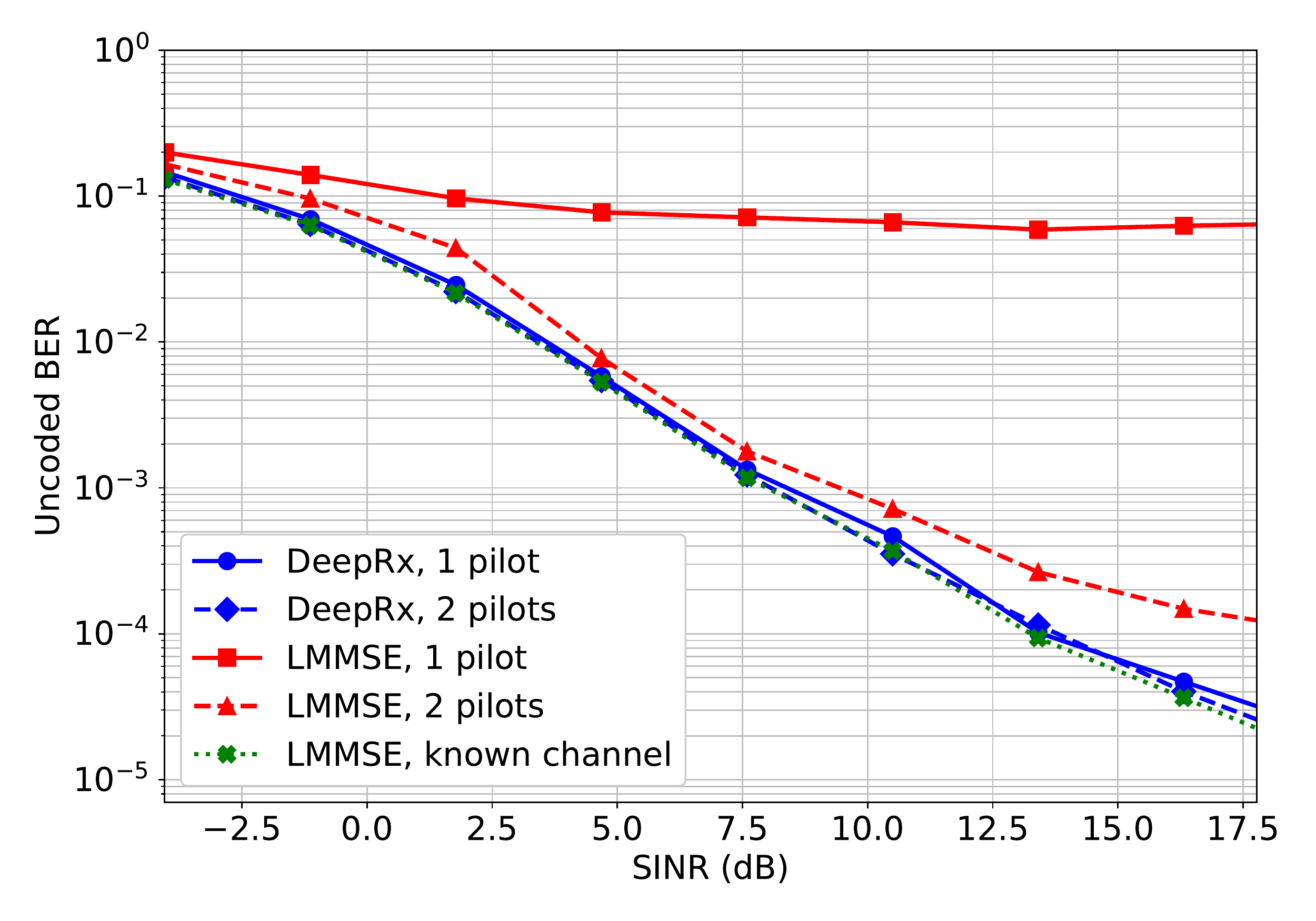} }}%
	\qquad
	\subfloat[QPSK data with manipulated transmit symbol distribution]{{\includegraphics[width=0.46\linewidth]
			{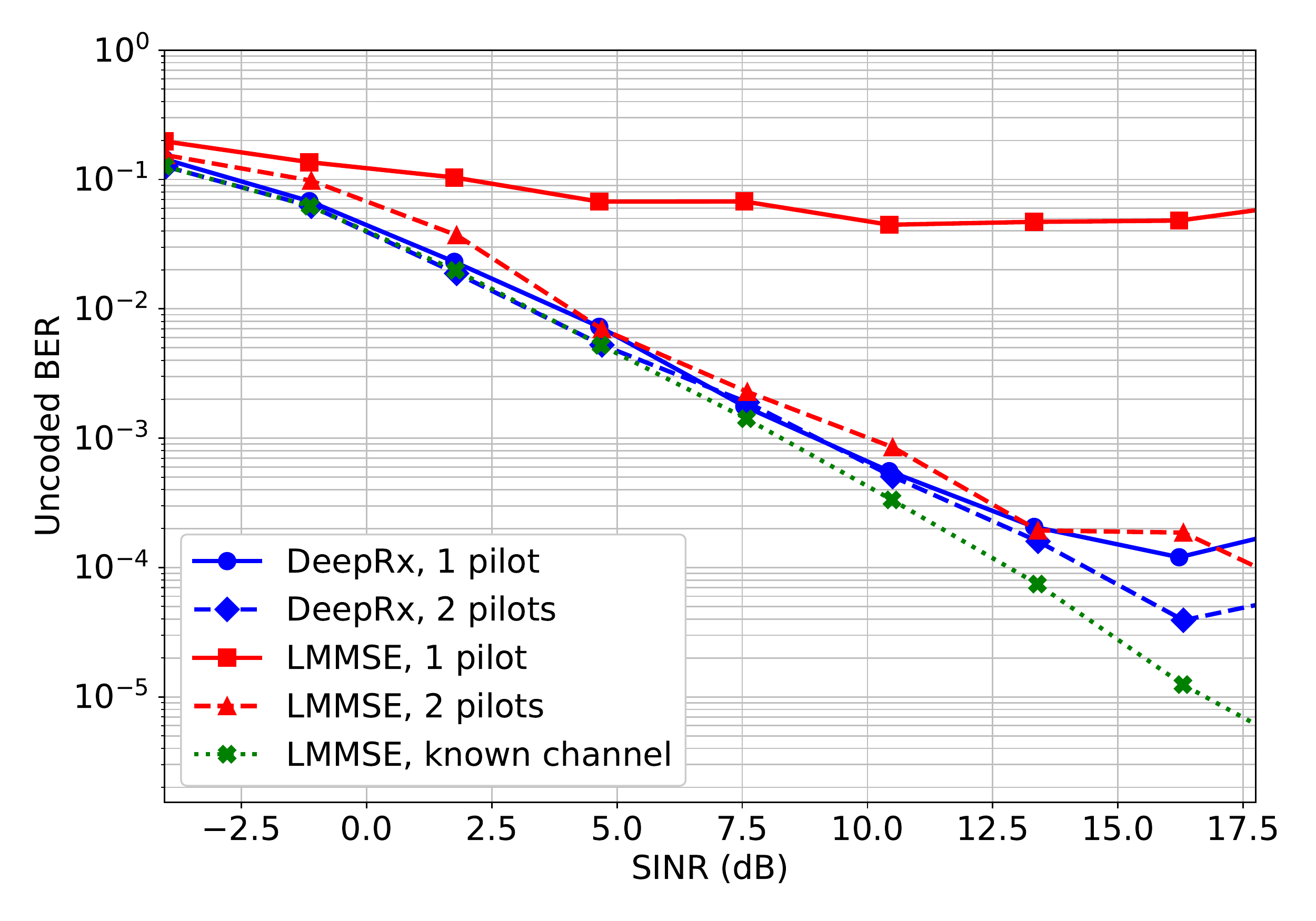} }}%
	\caption{(a) Uncoded BER after training and validating the DeepRx with regular QPSK data, and (b) uncoded BER after validating the DeepRx with QPSK data whose transmit symbol distribution is as depicted in Fig.~\ref{fig:corner_illustration}.}%
	\label{fig:corners}%
\end{figure}

\begin{figure*}[t!]
	\centering
	\includegraphics[width=0.45\textwidth]{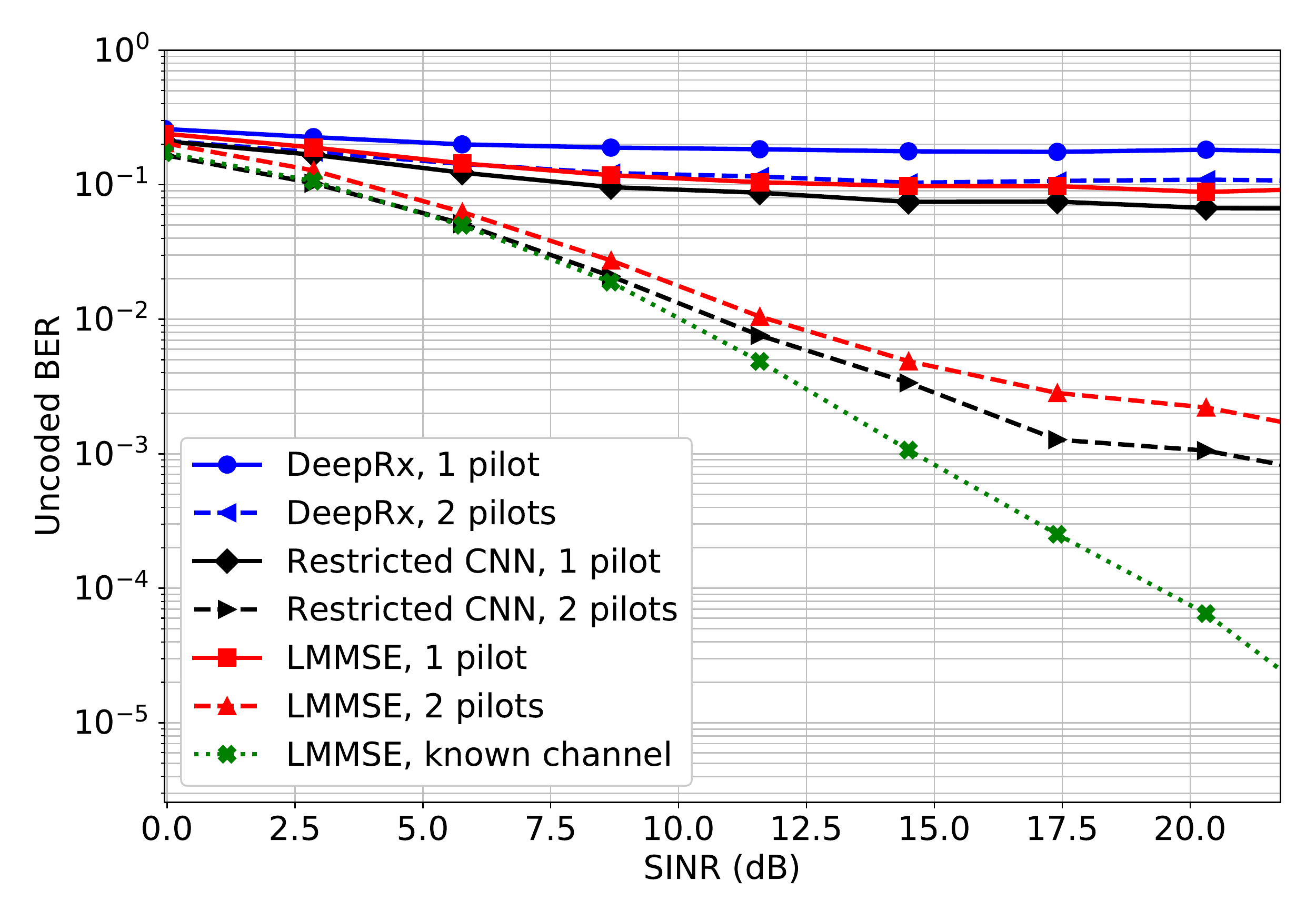}
	\caption{Uncoded BER of DeepRx and the restricted CNN receiver with 16-QAM data where transmit symbols are allocated as depicted in Fig.~\ref{fig:corner_illustration}. The models are the same that have been validated also in Figs.~\ref{fig:noint_allpilots} and~\ref{fig:restricted_nn}.} 
	\label{fig:corners_16qam}
\end{figure*}

\subsubsection*{Advanced Utilization of Data Symbol Distribution}

Another reason behind the high performance of DeepRx is that it might learn to utilize distributional information of data symbols (e.g., the known constellation points) to track the changes in the channel through time and frequency. For instance, DeepRx could learn to utilize the local distribution of symbols for blindly scaling and rotating the received data symbols to better match the known properties of the used constellation (compression and encryption of the transmitted data implicitly enforce a nearly uniform symbol distribution). To study this hypothesis, let us consider QPSK-modulated data in which each quadrant of the time-frequency grid of a TTI is filled with an identical symbol as shown in Fig.~\ref{fig:corner_illustration}, while still retaining the regular pilot symbol positions. We validate DeepRx\footnote{This version of DeepRx is trained with regular QPSK data with uniform random symbols. We did not include this particular type of artificial data (Fig.~\ref{fig:corner_illustration}) to the training data as it would lead to a task too easy to learn due to the known repetition of symbols.} with this type of data and compare the performance to a case where the validation data has a typical (nearly uniformly random) distribution of QPSK symbols. As shown in Fig.~\ref{fig:corners}, there is only a small performance reduction with this artificial validation data: DeepRx still clearly outperforms the practical LMMSE receiver, especially in the single-pilot case.

Next, we repeat the same experiment with 16-QAM constellation. Figure~\ref{fig:corners_16qam} shows results for a validation in which the data is constructed such that the time-frequency quadrants of each TTI are allocated with four randomly chosen 16-QAM symbols as shown in Fig.~\ref{fig:corner_illustration}.  Figure~\ref{fig:corners_16qam} also includes results for the restricted CNN receiver described above.  Now, when both magnitude and phase of a symbol are used to encode information (as opposed to QPSK where all the information is in the phase of the symbol), the results are very different. Namely, DeepRx fails to deliver sufficient performance under the manipulated transmit symbol distribution. Moreover, the restricted CNN receiver, which is not able to learn to utilize the data symbol distribution since it cannot observe multiple symbols at once, clearly outperforms the (unrestricted) DeepRx architecture.

These findings indicate that DeepRx learns to rely on the data symbol distribution to perform some type of local magnitude normalization or magnitude tracking to accurately equalize the channel amplitude response, and is now fooled when the distribution is artificially violated in the validation data. Indeed, the restricted CNN architecture, which can only process one symbol at a time, is not affected by this. This failure seems to be mostly related to the amplitude of the symbols, indicating that DeepRx might have learned some type of a blind equalization scheme that bears resemblance to the well-known constant-modulus algorithm (CMA) \cite{Godard80}. This becomes particularly evident when considering the accuracy of the individual bits detected by DeepRx, averaged over the whole SINR range. It is observed that the two most significant bits (corresponding to the phase of the symbol; see Fig. \ref{fig:constellations}a) remain very accurate (BER:~$9.0 \times 10^{-3}$), while the two least significant bits (requiring also magnitude information of the symbol) are very inaccurate under a manipulated data symbol distribution (BER:~$3.4 \times 10^{-1}$).

\subsubsection*{Comparison to Iterative Receiver Processing}
\label{sect:iterative}

Finally, we investigate further a hypothesis that DeepRx learns a reception technique which utilizes information about the legal constellation points. Such information is employed, for instance, in iterative receiver processing (see, e.g., \cite{Bonnet06a}). In order to gain further insight into this hypothesis, we created an additional data set where the channel is just a single uniformly distributed random phase shift for the whole TTI. This means that the channel for each TTI is essentially a scalar on the open interval of $(0,2 \pi )$. To make the task of channel estimation still sufficiently demanding, we use only a single pilot symbol located at the center subcarrier of the 3rd OFDM symbol in time\footnote{This is not to be confused with the one pilot cases presented in Fig.~\ref{fig:pilotconfig} and used in the other results, where the pilot covers one OFDM symbol but multiple subcarriers. In this particular experiment, each TTI has only one resource element allocated for a pilot symbol.}. For this experiment, we also implemented a simple iterative receiver, which is using its initial symbol decisions as additional information to further refine the channel estimate. More precisely, it performs the initial channel estimation using the single pilot symbol and equalizes the RX symbols, after which it calculates a new channel estimate using also the equalized RX symbols. This refined channel estimate is then averaged over the whole TTI, after which it can be used for equalizing the RX symbols again. This procedure is repeated 40 times to ensure that convergence is achieved.

DeepRx is executed with an otherwise identical configuration and architecture as earlier, except for increasing the sizes of the convolutional filters and lengths of dilations: all (3,3)-convolutional filters are changed to (10,3)-filters and dilations in the frequency direction are increased such that the maximum dilation 6 is increased to 16 and other dilations are increased with a similar ratio (cf. Table \ref{tab:cnn_arch}). This is done to ensure the visibility of the pilots as each TTI now contains just one pilot subcarrier, unlike in the primary scenario where the pilots span multiple subcarriers. Furthermore, we input two additional 2D arrays as channels to the network, one filled with the used frequencies and the other filled with the used time slots (both normalized to the unit interval). This allows the network to utilize time and frequency information, allowing the convolutional filters to specialize to this specific case of a single pilot\footnote{We wish to note that a fully convolutional architecture is not the optimal one for this simple task. The purpose of this experiment is merely to provide a simple setting for probing into the behavior of DeepRx, for which reason we did only minimal changes to its primary architecture.}.

\begin{figure*}[t!]
	\centering
	\includegraphics[width=0.46\textwidth]{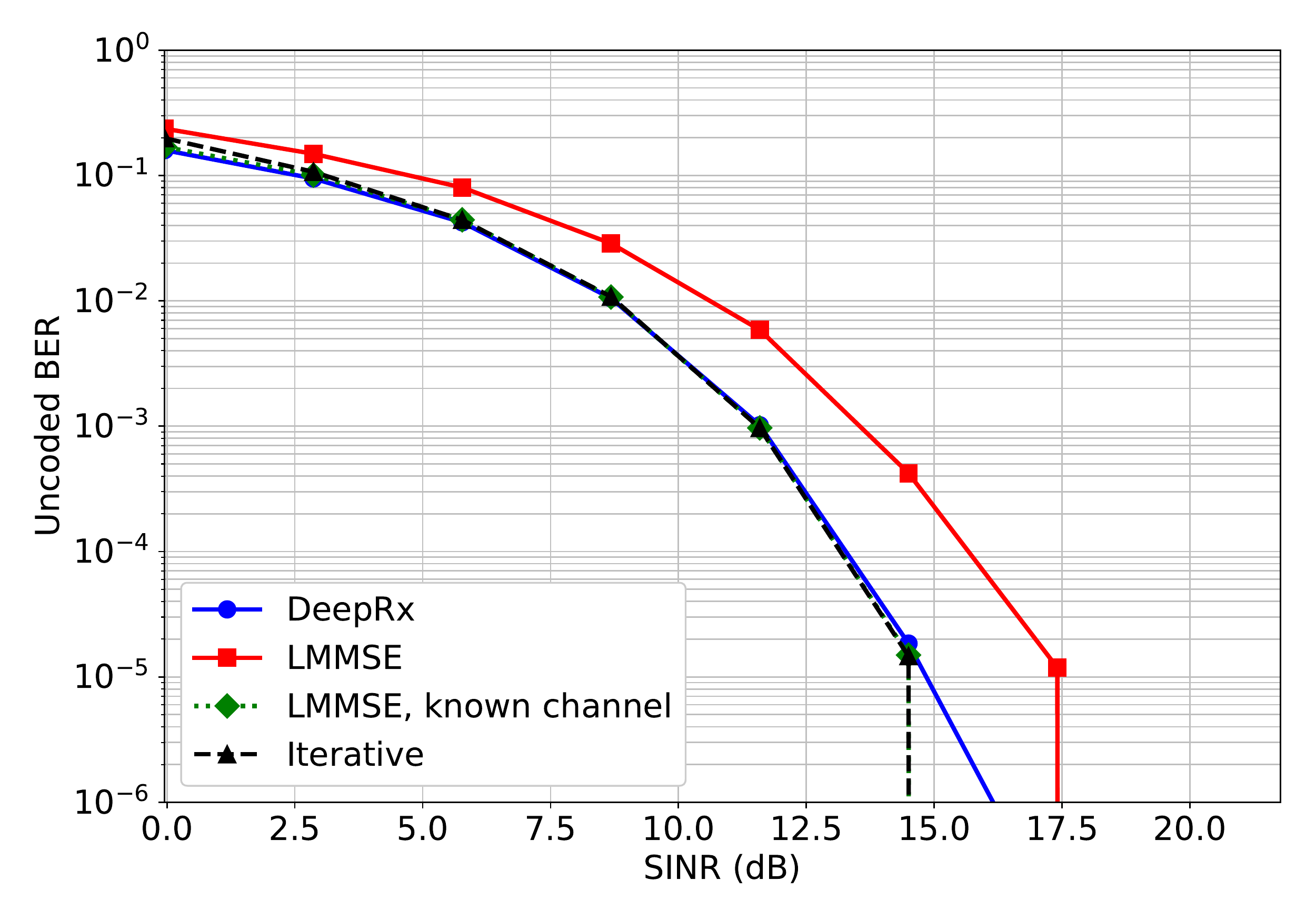}
	\caption{Uncoded BER after training and validating the DeepRx with data where the channel is just a random phase rotation.}
	\label{fig:phaserot_static}
	\vspace{-4mm}
\end{figure*}

The resulting uncoded BER performance for the different receivers is shown in Fig.~\ref{fig:phaserot_static}. Both the DeepRx and the iterative receiver are able to detect the channel almost perfectly, since they can utilize the unknown data symbols for channel estimation. On the other hand, the practical LMMSE receiver can only utilize a single pilot symbol for a noisy channel estimate, and thus falls short of the other solutions. This indicates that the processing learned by the DeepRx most likely resembles that of iterative receivers.

\begin{table}[!t]
	\renewcommand{\arraystretch}{1}
	\addtolength{\tabcolsep}{-1.5pt}
	\caption{Ablation study results (uncoded BER for uniformly distributed SINR between 15~dB and 20~dB). The architecture written in boldface refers to the primary DeepRx architecture (cf. Table \ref{tab:cnn_arch}) used to generate the results in Sections~\ref{sec:validation_res}-A and \ref{sec:validation_res}-B. The labels S, M, L and XL in the architecture name refer to small, medium, large and extra large in terms of number of parameters.} 
	\vspace{-0.1in}
	\label{tab:ablation}
	\centering
	\footnotesize
	
	\begin{tabular}{l|l|l|l|l|l|l|l|l}
		
		\textbf{Name} &   \textbf{\shortstack[l]{Depth, \\ ResNet \\ Blocks}} & \textbf{Params} & \textbf{Channels} & \textbf{\shortstack[l]{Min- \\ Max \\ Dilat.}} &  \textbf{\shortstack[l]{BER (1 pilot,\\interference)}} &   \textbf{\shortstack[l]{BER (2 pilots,\\interference)}} & \textbf{\shortstack[l]{BER (1 pilot,\\ no interference)}} &   \textbf{\shortstack[l]{BER (2 pilots,\\no interference)}}  \\ \hline\hline
		\multicolumn{8}{c}{\emph{Non-trainable baselines}}   \\ \hline
		\multicolumn{5}{l|}{LMSSE}&   $1.11 \times 10^{-1}$ & $3.38\times 10^{-3}$ &   $1.10 \times 10^{-1}$ &  $2.60 \times 10^{-3}$ \\ \hline
		\multicolumn{5}{l|}{LMSSE with known channel}&   $9.98 \times 10^{-4}$ &  $9.62 \times 10^{-4}$ & $4.51 \times 10^{-4}$ & $4.81 \times 10^{-4}$ \\ \hline
		\multicolumn{8}{c}{\emph{Large networks}}   \\ \hline
		11 XL 
		& 11 & 7M& 256-512 & 1-6 & $4.49 \times 10^{-4}$ &  $3.77 \times 10^{-4}$ \\ \hline
		11 L  
		& 11 & 3.4M& 128-512 & 1-6 & $4.95 \times 10^{-4}$  & $4.14 \times 10^{-4}$ \\ \hline
		13 L 
		& 13 & 1.8M& 64-256 & 1-6 &  $5.72 \times 10^{-4}$ & $4.71 \times 10^{-4}$\\ \hline
		\multicolumn{8}{c}{\emph{Different depths}}   \\ \hline
		13 M 
		& 13 & 1.2M& 64-128 & 1-6 & $6.19 \times 10^{-4}$  & $4.93 \times 10^{-4}$ & $4.37 \times 10^{-4}$ & $4.14 \times 10^{-4}$\\ \hline
		\textbf{DeepRx}& \textbf{11} & \textbf{1.2M} & \textbf{64-256} & \textbf{1-6} & $\bm{6.23 \times 10^{-4}}$  &$\bm{4.98 \times 10^{-4}}$ &  $\bm{4.47 \times 10^{-4}}$ &  $\bm{4.18 \times 10^{-4}}$\\ \hline
		5 M 
		& 5 & 1.2M& 192-256 & 1-6 & $1.09 \times 10^{-3}$  & $7.36 \times 10^{-4}$ &  $5.91 \times 10^{-4}$ &  $5.13 \times 10^{-4}$ \\ \hline
		3 M 
		& 3 & 1.2M& 256-448 & 1-6 &   $1.63 \times 10^{-3}$ &  $8.33 \times 10^{-4}$ \\ \hline
		\multicolumn{8}{c}{\emph{Depth multiplier (DM) $1$ instead of $2$ in all depthwise separable convolutions}}   \\ \hline
		11 S-DM1 & 11 & 0.6M & 64-128 & 1-6 &  & & $4.63 \times 10^{-4}$ & $4.24 \times 10^{-4}$\\ \hline
		\multicolumn{8}{c}{\emph{Different widths}}   \\ \hline
		11 S1 
		& 11 & 0.5M& 64-128 & 1-6 & $7.72 \times 10^{-4}$ &  $5.82 \times 10^{-4}$ \\ \hline
		11 S2 
		& 11 & 0.3M& 32-128 & 1-6 &  $8.63 \times 10^{-4}$ &  $6.32 \times 10^{-4}$ \\ \hline
		11 S3 
& 11 & 0.1M& 16-64 & 1-6 &  $3.72 \times 10^{-1}$  & $3.72 \times 10^{-1}$ \\ \hline
		11 S4 
		& 11 & 0.06M& 32 & 1-6 &  $2.58 \times 10^{-3}$  & $1.23 \times 10^{-3}$ \\ \hline
		\multicolumn{8}{c}{\emph{No dilation (ND)}}   \\ \hline
		11 M-ND  
		& 11 & 1.2M& 64-256 & 1 & $6.91 \times 10^{-4}$ &  $5.18 \times 10^{-4}$ \\ \hline
	    3 M-ND 
		& 3 & 1.2M& 256-448 & 1 & $1.14 \times 10^{-1}$  & $1.17 \times 10^{-3}$ \\ \hline
		\multicolumn{8}{c}{\emph{Regular convolutions, i.e., not depthwise separable (C)}}   \\ \hline
		11 M-C
		& 11 & 2.7M& 64-256 & 1-6 & $1.57 \times 10^{-3}$  & $8.71 \times 10^{-4}$ \\ \hline
		\multicolumn{8}{c}{\emph{Restricted CNN receiver, Fig.~\ref{fig:chan_est_diag} (R)}}   \\ \hline
		11 R & 11 & 1.2M& 64-256 & 1-6 &   &  &  $7.00 \times 10^{-2}$ & $1.29 \times 10^{-3}$\\ \hline
	\end{tabular}
\end{table}

\subsection{Ablation Studies and Notes on Complexity}
\label{sec:ablation}

Finally, to better understand the relationship between the exact architecture of DeepRx and its performance, we experimented with different CNN architectures and their hyperparameters. The results of this study are documented in Table \ref{tab:ablation}. The networks were trained with the full SINR range, while the BER values of the table are averages calculated for the SINR range from 15~dB to 20~dB for more informative results. The study was first performed using 16-QAM data with interference, after which we repeated it using the same hyperparameters and data without interference. Therefore the no-interference results can be considered as a separate test set. We also verified the lack of overfitting to the validation sets with a second test set generated with different random seed and confirmed that there was no significant difference between the results. 

For most of the runs, we have used the same optimizer hyperparameters defined earlier in Section \ref{sec:validation_res}. The notable exception is the model which uses normal convolutions instead of depthwise separable convolutions, where we had to divide the main learning rate by 2 in order for the model to converge.

Overall, it is clear from Table \ref{tab:ablation} that, with interference, almost all of the tested architectures outperform the LMMSE baselines. When tested with data without interference, the margin is smaller, but for cases with only one pilot, the performance gains from the CNN receivers are still considerable. For the remainder of this section, we consider the uncoded BER with one pilot and interference the primary benchmark metric for the different CNN architectures.

Regarding number of ResNet blocks (i.e., different depths), it seems that 11 blocks provide good performance under all scenarios, while still allowing for shrinking the network in terms of layer widths for more efficient inference. With dilations, it is possible to get relatively good performance with as few as 3 or 5 ResNet blocks, but the performance degrades slightly with one pilot and interference ($1.63 \times 10^{-3}$ for 3 blocks vs. $6.23 \times 10^{-4}$ for 11 blocks). In general, the dilation experiments indicate that the network requires a certain receptive field size to function well, although the deeper 11-block networks work well also without dilations ($6.91 \times 10^{-4}$ without dilations vs. $6.23 \times 10^{-4}$ with dilations).

Let us then investigate the effect of number of parameters, which is roughly proportional to the computational cost of inference for a given neural network (within certain conditions on the network architecture, such as number of layers and type of convolutions). Considering a CNN architecture of same depth of 11 ResNet blocks but with different widths, it is evident that roughly 1.2M parameters (primary DeepRx model with 64-256 convolutional channels, i.e., convolutional filters) provides a good balance between network complexity and BER performance. If only 0.1M parameters are used (model 11~S3 with 16-64 convolutional channels), the performance is heavily degraded. We also tried bigger networks, for example, model 11~XL with 7M parameters, and still observed some additional gains, but the performance increase started to level off after the 1M parameters mark. In order to explore smaller networks, we deviated from the general architecture a bit, and trained networks with constant width, and found a very small architecture (11~S4) with 32 convolutional channels, whose performance was relatively good ($2.58 \times 10^{-3}$, 1 pilot). By default, all of the above models utilizing depthwise separable convolution \cite{conf/cvpr/SifreM13,chollet2017} use a depth multiplier value of 2, with the exception of one model (11~S-DM1), which uses a depth multiplier of 1 in order to demonstrate that DeepRx is robust to changing this hyperparameter.

In addition, as mentioned previously, we tested a network without depthwise separable convolutions (model 11~M-C) and a restricted network that has to perform the channel estimation without having access to the unknown data symbols (model 11~R, similar to Fig.~\ref{fig:chan_est_diag}). As shown in Table~\ref{tab:ablation}, resorting to normal convolutions results in a slight drop in performance, while restricting the access to unknown data symbols deteriorated the performance to the level of the LMMSE baselines, as already observed in Section \ref{sec:expl-perf}.

As a final note, we emphasize that the computational complexity of DeepRx for scaling to larger TTIs is linear both in subcarrier and time dimensions (i.e., its asymptotic complexity can be expressed as $\mathcal{O}\left( S F \right)$), owing to the fully convolutional architecture. This applies to all the network architectures considered herein, as long as the pilot density within the TTI is kept roughly constant (otherwise receptive field might have to be adjusted to cover sparser pilots). On the other hand, the LMMSE receiver also has the complexity of $\mathcal{O}\left( S F \right)$ when making the reasonable assumption that its computational complexity is dominated by the equalization and demapping phase (see Section~\ref{sec:simulator}). This means that the theoretical asymptotic complexity of DeepRx in terms of the reception bandwidth and TTI length is similar to that of the LMMSE receiver, although DeepRx has a larger constant multiplier not visible in the asymptotic expressions. Therefore, we expect that in practice DeepRx has a higher computational cost than the LMMSE receiver, although it also achieves superior radio performance. As for scaling DeepRx to support a larger amounts of antennas, the impact on the computational complexity is more difficult to assess since it likely requires alterations to the network architecture such as increasing the number of convolutional filters for the first layers. Considering such multi-antenna aspects in more detail is left for future work.

\section{Conclusion}
\label{sec:conc}

In this paper we considered a ML-based digital radio receiver, trained as a supervised training task from frequency domain antenna signals into uncoded bits, and implemented in a 5G-compliant manner. Our hypothesis and primary motivation behind the work was that training the nearly complete digital receiver chain as a single supervised system would result in higher performance compared to training multiple smaller parts of the receiver separately. This allows for optimizing the system directly for the end task of recovering the transmitted bits. In addition, restricting the neural network architecture as little as possible allows it to learn improved, and potentially unforeseen, receiver schemes. With this, it could learn to implicitly solve various radio channel and hardware impairments which might otherwise be challenging to capture.

To address and investigate the hypothesis, we implemented a deep fully convolutional neural network, referred to as DeepRx, which was trained to detect the uncoded bits directly from the frequency-domain antenna signals. Moreover, DeepRx was trained to support different 5G-specific pilot configurations and modulation schemes. In contrast to many related works, the input of the neural network was constructed such that both the unknown data symbols as well as the known pilot symbols were arranged as convolutional input channels. This allowed DeepRx to efficiently combine both the data and pilot symbols when estimating the channel.

Through simulations modeling 5G uplink data transmission, we showed that the proposed DeepRx network outperforms traditional methods by a significant margin. It also outperformed an alternative neural network implementation where channel estimation and equalization were considered separately. We attributed the success primarily to DeepRx learning to utilize the known constellation points of the unknown data symbols, together with the local symbol distribution, to estimate and equalize the channel very accurately. In fact, some of the experiments indicated that the internal processing of DeepRx somewhat resembles that of iterative receivers. Moreover, it was also shown that DeepRx learns to deal efficiently with non-Gaussian interference and noise.

One restriction of this work is that it only includes a limited computational complexity analysis as the primary focus of this article is on the radio performance gains achieved by applying deep learning. Even though we have studied different approaches in improving the efficiency of the network, a further study on adapting these networks to inference time neural network chips is needed. Moreover, a fully comprehensive complexity comparison against conventional receivers is highly hardware-specific and needs to focus on latency and power consumption versus radio performance. While conducting such analysis is outside the scope of this article, it constitutes an important future work item for us. Another important future work topic is extending the DeepRx architecture to MIMO reception.

\section*{Acknowledgments}

We would especially like to thank Vesa Starck for outstanding support and excellent insights during this work. We would also thank Leo K\"arkk\"ainen, Mikko Uusitalo, Jakob Hoydis, and Andrew Baldwin for the various ideas and contributions. 

\bibliographystyle{IEEEtran}
\bibliography{IEEEabrv,references}

\end{document}